\documentclass[12pt,preprint]{aastex}
\usepackage{floatpag}

\shorttitle{Multiwavelength Observations of 4C $+$21.35}
\shortauthors{Ackermann et al.}

\begin{document}

\title{MULTIFREQUENCY STUDIES OF THE PECULIAR QUASAR 4C $+$21.35 DURING THE 2010 FLARING ACTIVITY}
\author{M.~Ackermann\altaffilmark{1}, 
M.~Ajello\altaffilmark{2}, 
A.~Allafort\altaffilmark{3}, 
E.~Antolini\altaffilmark{4,5}, 
G.~Barbiellini\altaffilmark{6,7}, 
D.~Bastieri\altaffilmark{8,9}, 
R.~Bellazzini\altaffilmark{10}, 
E.~Bissaldi\altaffilmark{11}, 
E.~Bonamente\altaffilmark{5,4}, 
J.~Bregeon\altaffilmark{10}
% ...
}
% ...
\altaffiltext{1}{Deutsches Elektronen Synchrotron DESY, D-15738 Zeuthen, Germany}
\altaffiltext{2}{Space Sciences Laboratory, 7 Gauss Way, University of California, Berkeley, CA 94720-7450, USA}
\altaffiltext{3}{W. W. Hansen Experimental Physics Laboratory, Kavli Institute for Particle Astrophysics and Cosmology, Department of Physics and SLAC National Accelerator Laboratory, Stanford University, Stanford, CA 94305, USA}
\altaffiltext{4}{Dipartimento di Fisica, Universit\`a degli Studi di Perugia, I-06123 Perugia, Italy}
\altaffiltext{5}{Istituto Nazionale di Fisica Nucleare, Sezione di Perugia, I-06123 Perugia, Italy}
\altaffiltext{6}{Istituto Nazionale di Fisica Nucleare, Sezione di Trieste, I-34127 Trieste, Italy}
\altaffiltext{7}{Dipartimento di Fisica, Universit\`a di Trieste, I-34127 Trieste, Italy}
\altaffiltext{8}{Istituto Nazionale di Fisica Nucleare, Sezione di Padova, I-35131 Padova, Italy}
\altaffiltext{9}{Dipartimento di Fisica e Astronomia ``G. Galilei'', Universit\`a di Padova, I-35131 Padova, Italy}
\altaffiltext{10}{Istituto Nazionale di Fisica Nucleare, Sezione di Pisa, I-56127 Pisa, Italy}
\altaffiltext{11}{Istituto Nazionale di Fisica Nucleare, Sezione di Trieste, and Universit\`a di Trieste, I-34127 Trieste, Italy}
% ...

\begin{abstract}

The discovery of rapidly variable Very High Energy (VHE; $E >$ 100 GeV) $\gamma$-ray emission from
4C $+$21.35 (PKS\, 1222+216) by MAGIC on 2010 June 17, triggered by the high activity detected by the {\em Fermi} Large Area
Telescope (LAT) in high energy (HE; $E >$ 100 MeV) $\gamma$-rays, poses intriguing
questions on the location of the $\gamma$-ray emitting region in this
flat spectrum radio quasar.  We present multifrequency data of 4C $+$21.35 collected from centimeter to VHE during 2010 to investigate the properties of this
source and discuss a possible emission model. The first hint of detection at
VHE was observed by MAGIC on 2010 May 3, soon after a $\gamma$-ray flare detected by {\em Fermi}-LAT that peaked on April 29.  The same
emission mechanism may therefore be responsible for both the HE and
VHE emission during the 2010 flaring episodes. Two optical peaks were
detected on 2010 April 20 and June 30, close
in time but not simultaneous with the two $\gamma$-ray peaks, while no
clear connection was observed between the X-ray and $\gamma$-ray
emission. An increasing flux density was observed in radio and mm
bands from the beginning of 2009, in accordance with the increasing
$\gamma$-ray activity observed by {\em Fermi}-LAT, and peaking on 2011
January 27 in the mm regime (230 GHz).  We model the spectral energy
distributions (SEDs) of 4C $+$21.35 for the two periods of the VHE
detection and a quiescent state, using a one-zone model with the emission
  coming from a very compact region outside the broad line region.  The three
SEDs can be fit with a combination of synchrotron self-Compton and external Compton emission of seed photons from a dust torus, changing
only the electron distribution parameters between the epochs. The fit of
the optical/UV part of the spectrum for 2010 April 29 seems to favor an inner
disk radius of $<$6 gravitational radii, as one would expect from a prograde-rotating Kerr black hole.

\end{abstract}


\section{Introduction}
\label{intro}

In the last few years flat spectrum radio quasars (FSRQs) have been
established as a distinct Very High Energy (VHE) $\gamma$-ray blazar subclass. So far three FSRQs
have been detected at $E$ $>$ 100 GeV: 3C 279 \citep{albert08}, 4C
$+$21.35 \citep{MAGIC_discovery}, and PKS\, 1510$-$089
\citep{cortina12,abramowicz13}. These detections were surprising. The VHE
emission from FSRQs is expected to be absorbed internally, if emitted
within the broad line region (BLR), or externally, for sources located
at redshifts where the emission is strongly attenuated by
$\gamma$$\gamma$ pair production via interaction with the Extragalactic Background
Light (EBL) photons. In addition, since FSRQs usually have their
synchrotron peak at relatively low frequencies (i.e., infrared/optical
bands rather than UV/X-ray), their corresponding inverse Compton peak should fall at photon energies less than 1 GeV,
and thus a detection at VHE is not expected.

The FSRQ 4C $+$21.35 (also known as PKS\, 1222$+$21) has a redshift of  $z = 0.432$ \citep{osterbrock87} with a peculiar bent large-scale
radio structure \citep{saikia93}. Very large apparent superluminal motion
($\beta_{\rm app} \sim$ 10--15) has been detected on
milliarcsecond scales for sub-components of the jet \citep{jorstad01,
  homan01}. On the other hand, the ratio of the core-to-extended radio
fluxes at GHz frequencies is of the order of unity; thus it is
formally a ``lobe-dominated quasar'' \citep{kharb04, wang04}.

In GeV $\gamma$-rays the source was in a quiescent state from the
start of the {\em Fermi Gamma-ray Space Telescope} mission in 2008
August until 2009 September. After this period a gradually increasing flux
was observed, up to an interval of flaring activity in the first
half of 2010. In particular, 4C $+$21.35 underwent two very strong
outbursts in 2010 April and June, observed by the Large Area Telescope
(LAT) on board {\em Fermi} and composed of several major flares
characterized by rise and decay timescales of the order of one day
\citep{tanaka11}.  During the second flaring activity, VHE emission
from 4C $+$21.35 was detected with the MAGIC Cherenkov telescopes on
2010 June 17, with a flux doubling time of about 10 minutes \citep{MAGIC_discovery}. The simultaneous {\em Fermi}-LAT and Major
Atmospheric Gamma Imaging Cherenkov (MAGIC) spectrum, corrected for EBL absorption, can be described by a single power-law with photon index
$\Gamma_{\gamma}$ = 2.72 $\pm$ 0.34 between 3 GeV and 400 GeV, consistent with
emission from a single component in the jet. The absence of a spectral
cut-off for $E$ $<$ 130 GeV constrains the $\gamma$-ray emission
region to lie outside the BLR, which would otherwise absorb the $\approx$ 10-20 GeV photons by $\gamma\gamma\rightarrow e^\pm$ production when these $\gamma$-rays pass through the intense circum-nuclear photon fields provided by the BLR itself.
At the same time, the rapid VHE variability observed suggests an extremely compact
emission region, with size $R \le c t_{\rm var} \delta_D/(1+z) \sim 10^{15}\ (\delta_D/80)\ (t_{\rm var}/10\ {\rm minutes})$\ cm where $t_{\rm var}$ is
  the variability timescale and $\delta_D$ is the Doppler factor.  If the blob
  takes up the entire cross section of the jet, it implies that the emitting region is at a distance $r \sim
R/\theta_{\rm open} \sim 5.7\times10^{16} (\delta_D/80)\ (t_{\rm var}/10\ {\rm minutes})$\ ($\theta_{\rm open}/1\deg)^{-1}$\ cm, where $\theta_{\rm open}$ is
the half-opening angle of the jet.  Even for a highly relativistic jet with
$\delta_D \sim 100$, the location of the emission region should be well within
the BLR radius for 4C +21.35, likely $R_{\rm BLR}\approx 2\times10^{17}$\ cm \citep{tanaka11}.

Different models have been proposed to explain the unusual behavior of
4C $+$21.35. A very narrow jet can preserve
variability at the pc scale, but the likelihood of being in the beam
of such a thin jet is small, unless there were many narrow jets, as in
a jets-within-jet/mini-jets scenario
\citep{gt08,gub09,tav11}. An alternative model is a compact emission region at
the pc scale responsible for the emission at higher energies, with a second zone either inside
or outside the BLR to complete the modeling of the emission at lower energies \citep{tav11}. The compact emission sites at
the pc scale could be due to self-collimating jet structures
\citep{nal12}, where the magnetic field dominates the energy density, or to turbulent cells
\citep[e.g.,][]{nal11,mj10}. Another possibility is that the
acceleration of ultra-high energy cosmic rays protons in the inner jet leads
to an outflowing beam of neutrons that deposit their energy into
ultra-relativistic pairs that radiate VHE synchrotron emission at the
pc scale \citep{dmt12}, with associated neutrino production. Even more
  exotic scenarios have been proposed, such as photons produced inside the
  BLR, tunneling through it via photon to axion-like particle oscillations \citep{tavecchio12}.

In this paper, we present the multifrequency data of 4C $+$21.35 collected from
radio to VHE during 2010, and discuss a possible emission model for
this source. A summary of the complete multiwavelength data of 4C
$+$21.35 presented in this paper and the relative facilities can be found in Table \ref{facilities}.
The paper is organized as follows: in Sections 2 and 3 we briefly
report the LAT and MAGIC data analysis and results, respectively. In Section 4 we report the
result of {\em Swift} optical/UV/X-ray observations. Optical data collected by
the Abastumani, ATOM, Catalina, Crimean, KVA, Steward, and St. Petersburg observatories are presented in Section 5. In Section 6
we present the radio and mm data collected by the Medicina, UMRAO, MOJAVE,
OVRO, F-GAMMA, Mets\"ahovi, and SMA facilities. In Section 7 we discuss the
light curves behavior and the spectral energy distribution (SED) modeling of
three different epochs, and finally we draw our conclusions in Section 8.

Throughout the paper, a $\Lambda$ CDM cosmology with $H_0$ = 71 km s$^{-1}$ Mpc$^{-1}$, $\Omega_{\Lambda} = 0.73$, and $\Omega_{m} =
0.27$ is adopted. The corresponding luminosity distance at $z = 0.432$ is $d_L = 2370$\ Mpc, and 1 arcsec corresponds to a projected size of 5.6 kpc. 

\begin{table}[!hhh]
\caption{Observatories Contributing to the Presented Data Set of 4C $+$21.35 at Different Frequencies.}
\begin{center}
\begin{tabular}{ccc}
\hline \hline
\label{facilities}
Waveband & Observatory  & Frequency/Band  \\ 
\hline
Radio & SMA & 230 GHz \\
& Mets\"ahovi & 37 GHz \\
& VLBA (MOJAVE) & 15 GHz \\
& OVRO & 15 GHz \\
& UMRAO & 8.0, 14.5 GHz  \\
& Medicina & 5, 8 GHz \\
& F-GAMMA & 2.6, 4.8, 8.4, 10.5, 14.6, 23.1, 32, 86.2, 142.3 GHz \\
\hline
Optical & Abastumani & $R$ \\
& ATOM & $R$ \\
& Catalina & $V$ \\
& Crimean & $R$ \\
& KVA & $R$ \\
& St. Petersburg & $R$ \\ 
& Steward & $V$ \\
& {\em Swift}-UVOT & $v$, $b$, $u$ \\
\hline
UV & {\em Swift}-UVOT & $w1$, $m2$, $w2$  \\
\hline
X-rays & {\em Swift}-XRT & 0.3--10 keV \\
& {\em Swift}-BAT & 15--50 keV \\
\hline
HE $\gamma$-rays & {\em Fermi}-LAT & 0.1--300 GeV \\
VHE $\gamma$-rays & MAGIC & 70 GeV--5 TeV \\
\hline \hline
\end{tabular}
\end{center}
\end{table}

\section{MAGIC Data and Analysis}

The MAGIC experiment is situated in the Observatorio del Roque de los Muchachos in the Canary Island of La
Palma ($28^{\circ}45'$ north, $18^{\circ}54'$ west), 2200 \,m above sea
level. It consists of two 17-m Imaging Atmospheric Cherenkov Telescopes and can reach an energy threshold as low as 50\,GeV in standard trigger mode. Details on the performance of the MAGIC telescope stereo system can be found in \citet{MAGIC_performance}.

MAGIC observed 4C $+$21.35 between 2010 May 3 and June 19. In total, 16 hr of good quality data were collected. The data analysis was performed in
the MAGIC Analysis and Reconstruction Software analysis framework \citep{MARS,lombardi11}. On May 3 (MJD 55319), MAGIC obtained an excess
  with respect to the background of $\approx$ 78
events in 2.2 hr of observation, which yielded a marginal detection with a
signal significance of 4.4\,$\sigma$ using the Equation (17) of \citet{LiMa}. On June
17 (MJD 55364), MAGIC obtained a $\gamma$-ray excess of 190 events in a 30-minute long
observation, yielding a signal significance of 10.2 $\sigma$, implying the
first significant detection of this source in VHE $\gamma$-rays
\citep{MAGIC_discovery}. The VHE detection on June 17 shows fast variability with a flux doubling time of $8.6^{+1.1}_{-0.9}$ minutes, which is the fastest time
variation ever observed in a FSRQ, and among the shortest time scales measured
for TeV emitters \citep[see, e.g.,][]{albert07,aharonian07}. The observed
spectrum covered the energy range from 70\,GeV up
to at least 400\,GeV and can be fit with a single power-law with photon
index $\Gamma_{\gamma} = 3.75 \pm 0.27$. The spectrum corrected for the effect of EBL absorption making use of the EBL model
from \cite{dominguez11} can be also described by a single power-law with
photon index $\Gamma_{\gamma} = 2.72 \pm 0.34$ between 3 GeV and 400 GeV
  \citep[see][]{MAGIC_discovery}.
 
None of the other nights showed a statistically significant excess of signal over the
background. Upper limits at 95\% C.L. were calculated above 150 GeV assuming a
power-law with the same photon index measured on June 17 (i.e. $\Gamma_{\gamma}$ = 3.75) for the nights between May 5
and June 13. The rest of the nights were not included in the upper limit
calculation due to strong moonlight constraints. The upper limits range between 1.4\% Crab units (C.U.) (on May 30; MJD
55346) and 12.7\% C.U. (on June 12; MJD 55359), as reported
  in Table~\ref{MAGIC_UL}. Considering the period from May 5 to June 13
(total time: 12.5 hr) an upper limit of 1.6\% C.U. was estimated.

\begin{table*}
\caption{Daily Upper Limits Estimated by MAGIC for $E>$150 GeV Assuming a Spectrum Slope 3.7}
\begin{center}
\begin{tabular}{cccc}
\hline \hline
\multicolumn{1}{c}{Date} &
\multicolumn{1}{c}{Effective Time} &
\multicolumn{1}{c}{Integral Limit} &
\multicolumn{1}{c}{Integral Limit} \\
\multicolumn{1}{c}{(UT)} &
\multicolumn{1}{c}{(hr)} &
\multicolumn{1}{c}{(cm$^{-2}$ s$^{-1}$) above 150 GeV} &
\multicolumn{1}{c}{(in MAGIC C.U.)} \\
\hline
2010 May 5 & 0.5 & 1.2e-11 & 3.7\% \\
2010 May 6 & 0.7 & 1.1e-11 & 3.3\% \\
2010 May 7 & 0.8 & 1.8e-11 & 5.5\% \\
2010 May 8 & 1.4 & 1.7e-11 & 5.4\% \\
2010 May 30 & 0.9 & 4.6e-12 & 1.4\% \\
2010 May 31 & 1.0 & 2.6e-11 & 8.1\% \\
2010 June 1 & 1.2 & 5.1e-12 & 1.6\% \\
2010 June 2 & 0.9 & 1.0e-11 & 3.7\% \\
2010 June 3 & 1.1 & 8.4e-12 & 2.6\% \\
2010 June 4 & 1.2 & 8.0e-12 & 2.5\% \\
2010 June 6 & 1.0 & 1.4e-11 & 4.3\% \\
2010 June 7 & 0.6 & 2.1e-11 & 6.4\% \\
2010 June 8 & 0.7 & 1.3e-11 & 4.0\% \\
2010 June 9 & 0.9 & 2.4e-11 & 7.3\% \\
2010 June 12 & 0.6 & 4.1e-11 & 12.7\% \\
2010 June 13 & 0.6 & 2.5e-11 & 7.8\% \\
\hline
\hline
\end{tabular}
\end{center}
\label{MAGIC_UL}
\end{table*}

\section{{\em Fermi}-LAT}

The {\em Fermi}-LAT is a $\gamma$-ray telescope operating from $20$\,MeV to
above 300 GeV. The LAT has a large peak effective area ($\sim$ $8000$\,cm$^2$
for $1$\,GeV photons), a relative energy resolution typically $\sim$10\%, and a field of
view of about $ 2.4$ \,sr with an angular resolution ($68\%$ containment angle) better than 1\degr\ for
energies above $1$\,GeV. Further details about the LAT are given by
\citet{atwood09}. 

Very strong GeV flares from 4C $+$21.35 were detected by {\em
Fermi}-LAT in 2010 April and June, with spectra characterized by a
broken power-law with spectral breaks near 1-3 GeV photon
energies and a photon index after the break $\sim$2.4. In contrast, the
quiescent state observed by the LAT during 2008 August--2009 September has
been fit by a single power-law with photon index $\Gamma_{\gamma}$ = 2.57
$\pm$ 0.07. 
All details of the LAT analysis for that period were presented in \citet{tanaka11}. 
After the 2010 flaring period, a decreasing $\gamma$-ray activity of 4C
$+$21.35 was observed by {\em Fermi}-LAT, and then in mid-2011 the source
faded back into a quiescent state.\footnote{{\footnotesize
    http://fermi.gsfc.nasa.gov/FTP/glast/data/lat/catalogs/asp/current/lightcurves/PKSB1222+216\_86400.png}}
4C $+$21.35 is found in the first {\em Fermi} hard source list (1FHL) as 1FHL
J1224.8$+$2122 \citep{ackermann13}. This object is the most variable
  source in the 1FHL catalog. 

\section{{\em Swift} Observations}

\label{SwiftData}

The {\em Swift} satellite \citep{gehrels04} performed 28 observations of 4C
$+$21.35 between 2010 February 12 and June 23. The observations were performed
with all three on-board instruments: the Burst Alert Telescope \citep[BAT;][15--150 keV]{barthelmy05}, the X-ray Telescope \citep[XRT;][0.2--10.0
  keV]{burrows05}, and the UltraViolet Optical Telescope \citep[UVOT;][170--600 nm]{roming05}. 

\subsection{{\em Swift}/BAT}

4C $+$21.35 is detected in the BAT 70-month catalog, generated from the all-sky survey in the time period 2004 November--2010 August. The data
reduction and extraction procedure of the 8-channel spectrum is described in \citet{baumgartner13}. The 14--195 keV spectrum is well described by a
power-law with photon index of 1.76$_{-0.23}^{+0.25}$ ($\chi$$^2_{\rm red}$ =
0.60, 6 d.o.f.). The resulting unabsorbed 14--195 keV flux is
(2.2$\pm$0.4)$\times$10$^{-11}$ erg cm$^{-2}$ s$^{-1}$. No significant variability was observed in the BAT light curve on monthly time scales. Nevertheless, the hard X-ray flux of this source is below the sensitivity of the BAT instrument for the short exposure times of single {\em Swift} observations.

\subsection{{\em Swift}/XRT}

The XRT data were processed with standard procedures (\texttt{xrtpipeline v0.12.6}), filtering, and screening criteria by using the \texttt{HEASoft} package
(v6.11). The data were collected in photon counting mode in all observations, and only XRT event grades 0--12 were selected. The source count rate
was low ($<$ 0.5 counts s$^{-1}$), thus pile-up correction was not
required. Data collected in the same day were summed in order to have better
statistics. Since the observation performed on 2010 May 26 was short ($\sim$500 s), it was not considered. Source events were extracted from a circular region with a radius of
20 pixels (1 pixel = 2\farcs36), while background events were extracted from
a circular region with radius of 50 pixels away from the source region. Ancillary response files were generated
with \texttt{xrtmkarf}, and account for different extraction regions,
vignetting and point-spread-function corrections. When the number of photons collected was smaller
than 200 the Cash statistic was used \citep{cash79}. 

We fit the spectra for all the individual {\em Swift} observations with an absorbed power-law with a neutral hydrogen column
density fixed to its Galactic value \citep[$N_{\rm H}$= 2.09$\times$10$^{20}$ cm$^{-2}$; ][]{kalberla05}. The X-ray light curve and spectral shape derived from these fits is discussed in Section \ref{lc} together with the other multiwavelength data.

\subsection{{\em Swift}/UVOT}

The script that handles the UVOT analysis is \verb+uvotgrblc+ (available
within \texttt{HEASoft}). It determines the aperture corrected magnitude by (1)
selecting the aperture size based on the observed source count-rate and the
presence of close field sources; (2) choosing the background region based on
the surface brightness among three annular regions centered on the main source in
the summed images (circular regions around field sources are  automatically excluded); (3) finding field stars to estimate the aperture correction, specific for
each observation; (4) calling the task \verb+uvotsource+ to estimate the photometry.

Since 4C +21.35 is a very bright object in the optical and UV range and lies
in a sparsely populated area of the sky, \verb+uvotgrblc+ selected a
circle of 5\arcsec\ as the source extraction region and a full annulus for the
background region for all the observations. The typical inner/outer radii for
the background regions were 27\arcsec/35\arcsec\ and 35\arcsec/42\arcsec. The
UVOT magnitudes during these observations showed ranges as follows: $v$ = 15.67--15.21, $b$ =
  15.65--15.43, $u$ = 14.67--14.34, $w1$ = 14.37--14.08, $m2$ = 14.25--13.90,
  $w2$ = 14.16--13.90, with a typical error of 0.06 mag. As discussed in
  detail in Section \ref{lc}, no significant increase in flux was observed by UVOT during 2010 February--June, but the sparse
coverage does not allow us to draw firm conclusions.
 
\section{Optical Observations}

In this section we briefly describe the programs performing optical observations of 4C $+$21.35 and the corresponding data analysis. These optical data are discussed together with the multiwavelength data in Section \ref{lc}.

\subsection{Abastumani, Crimean and St. Petersburg Data}

Observational data at optical wavelengths ($R$-band) were obtained  at the 0.7-m reflector of the Crimean Astrophysical Observatory and 0.4-m LX-200 telescope of the Astronomical Institute of St. Petersburg State University, both equipped with photo-polarimeters  based on  ST-7XME CCDs. A standard technique of bias and dark subtraction and flat-field correction was used. Photometric calibration was made relative to two nearby standard stars, located in the same field.

Optical observations in $R$-band were performed also by the 0.7-m meniscus f/3 telescope of Abastumani Astrophysical Observatory in Abastumani, Georgia. 

\subsection{ATOM Data}

Optical observations in $R$ filter for this campaign were obtained
between 2010 February and May with the 0.8-m optical telescope ATOM in Namibia
\citep{hauser04}. ATOM is operated robotically by the H.E.S.S. collaboration
and obtains automatic observations of confirmed or potential $\gamma$-bright
blazars. Data analysis \citep[debiassing, flat fielding, and photometry with
  Source-Extractor;][]{bertin96} is conducted automatically using the pipeline developed by the ATOM Team.

\subsection{Catalina Real-Time Transient Survey}

The source is monitored by the Catalina Real-Time Transient Survey
\citep[CRTS\footnote{http://crts.caltech.edu};][]{drake09, djorgovski11},
using the 0.68-m Schmidt telescope at Catalina Station, AZ, and an unfiltered
CCD.  The typical cadence is to obtain four exposures separated by 10 minutes in a
given night; this may be repeated up to four times per lunation, over a period of $\sim$6--7 months each year, while the field is
observable.  Photometry is obtained using the standard Source-Extractor
package \citep{bertin96}, and roughly calibrated to the $V$-band in terms of
the magnitude zero-point. The light curve, accessible through the CRTS Web site and
spanning $\sim$6 yr, shows a dramatic increase in optical variability of
this source starting in late 2009. 

\subsection{KVA Data}

The KVA (Kungliga Vetenskaps Akademientelescope) is located on Roque de los Muchachos, La Palma (Canary Islands),
and is operated by the Tuorla Observatory, Finland
(http://users.utu.fi/kani/1m). The telescope consists of a 0.6-m f/15 Cassegrain devoted to polarimetry, and a
0.35-m f/11 SCT auxiliary telescope for multicolor photometry. The telescope has been successfully operated remotely since autumn 2003. The KVA is used for
optical support observations for MAGIC by making $R$-band photometric
observations, typically one measurement per night per source. The data were
reduced by the Tuorla Observatory Team as described in K. Nilsson et al. (2014, in preparation).

\begin{figure}
\centering
\includegraphics[width=11.0cm]{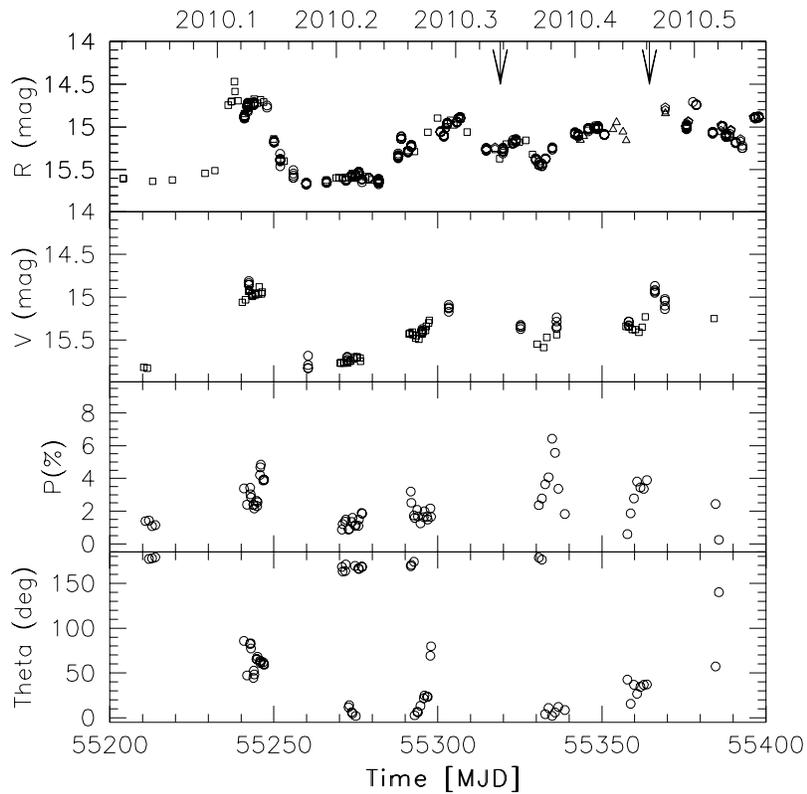}
\caption{Optical magnitudes in the $R$-band (top panel; circles: Abastumani, squares: ATOM, triangles: KVA, pentagons: Crimean and St. Petersburg), $V$-band (second panel; circles: CRTS, squares: Steward), percentage of polarized flux (third panel) and polarization position angle in $V$-band (bottom panel) are shown. For clarity the errors (typically $<$5\%) are not shown. The downward arrows indicate the times of the VHE detections by MAGIC.}
\label{polla}
\end{figure}

\subsection{Steward Observatory}\label{polar}

4C $+$21.35 was systematically monitored by Steward Observatory
during the {\em Fermi} observations, providing spectropolarimetry and
spectrophotometry measurements of this object in $V$-band,\footnote{http://james.as.arizona.edu/$\sim$psmith/Fermi/} as reported also in
\citet{smith11}. Figure \ref{polla} shows the behavior of the observed
degree of optical polarization P, and the position
angle of the polarization vector $\Theta$ as a function of
time. Visual inspection of the plot seems to show that in general
periods of high flux correspond to periods of relatively high
polarization degree and only small rotation of the polarization angle
  vector. In particular, a marginal increase of polarization degree but no significant rotation of the polarization angle was observed in 2010 June, during the period of HE and VHE flaring activity.

Spectrophotometry during 2010--2011 did not show significant changes in
the flux of the broad Mg II $\lambda$2800 and Balmer emission lines despite large
optical continuum variation. This indicates that non-thermal emission
from the jet has a negligible influence on the BLR
lines fluxes. \citet{smith11} suggested that the beaming jet emission
intersects only a small fraction of the volume containing the
emission-line gas. A different behavior was observed in 3C 454.3, with a
  significant increase of the Mg II emission line flux during the
  optical and $\gamma$-ray flaring activity in 2010 November. This event
  occurred after a mm flare onset, during an increase in the optical
  polarization percentage, and before the emergence of a superluminal knot
  from the radio core. This suggests the presence of BLR clouds surrounding
  the radio core in 3C\,454.3 \citep{leon13}.

\section{Radio and mm observations}

In this section we present the radio and mm light curves and spectra of 4C
$+$21.35 collected between 2009 January 1 and 2011 February 28 to investigate
their connection with the $\gamma$-ray activity. The data collected
  between 230 GHz and 5 GHz are reported in Figure \ref{radio} and discussed in
  detail in Section \ref{lc}. In addition we studied the radio structure and jet kinematics of this source during the
  MOJAVE monitoring observations.

\subsection{SMA Data}

The 230 GHz (1.3 mm) light curve was obtained at the Submillimeter Array (SMA)
on Mauna Kea (Hawaii). 4C $+$21.35 is included in an ongoing
monitoring program at the SMA to determine the fluxes of compact extragalactic
radio sources that can be used as  calibrators at mm wavelengths. Details of the observations and data reduction
can be found in \citet{gurwell07}. Data from this program are updated regularly and are available at the SMA Web site.\footnote{http://sma1.sma.hawaii.edu/callist/callist.html. Use of SMA data in publication requires obtaining permission in advance.}

\subsection{F-GAMMA Project}\label{effelsberg}

Radio spectra and light curves  of 4C $+$21.35 were obtained within the framework of a
{\it Fermi}-related monitoring program of $\gamma$-ray blazars \citep[F-GAMMA
project;][]{fuhrmann07}. The frequency range spans  2.64 GHz to 142 GHz
using the Effelsberg 100-m and IRAM 30-m telescopes. The Effelsberg measurements were conducted with the
secondary focus heterodyne receivers at 2.64, 4.85, 8.35, 10.45, 14.60, 23.05,
32.00, and 43 GHz. The observations were performed quasi-simultaneously with cross-scans, that is,
slewing over the source position, in azimuth and elevation directions, with
adaptive numbers of sub-scans for reaching the desired sensitivity \citep[for
details, see][]{fuhrmann08, angelakis08}. Pointing offset correction, gain
correction, atmospheric opacity correction, and sensitivity correction have
been applied to the data. The IRAM 30-m observations were carried out with
calibrated cross-scans using the EMIR horizontal and vertical polarization
receivers operating at 86.2 and 142.3 GHz. The opacity-corrected intensities were converted into the
standard temperature scale and finally corrected for small remaining pointing offsets and systematic gain-elevation effects.
The conversion to the standard flux density scale was done using the instantaneous conversion factors derived from frequently
observed primary (Mars, Uranus) and secondary (W3(OH), K350A, NGC 7027)
calibrators. The radio spectra from 2.64 GHz to 43 GHz obtained during five epochs of Effelsberg 
observations between 2009 January 24 and 2011 April 29 are shown in
Figure~\ref{Eff}. A significant increase of the flux density has been observed
from 2009 May to September at 43 GHz, while at longer wavelengths the increase
occurs later, likely due to synchrotron self-absorption opacity effects. This time difference led to a
significant radio spectral evolution, possibly related to the activity observed in $\gamma$-rays. 

\subsection{Mets\"ahovi Data}

The 37 GHz observations were made with the 13.7-m diameter Mets\"ahovi radio
telescope, which is a radome enclosed paraboloid antenna situated in
Finland (24 23' 38''E, +60 13' 05''). The measurements were made with a
1 GHz-band dual beam receiver centered at 36.8 GHz. The HEMPT (high electron
mobility pseudomorphic transistor) front end operates at room temperature.
The observations were taken with an ON--ON technique, alternating the source and the
sky in each feed horn. A typical integration time to obtain one flux
density data point is between 1200 and 1400 s. The detection limit of the telescope at 37 GHz is on the order of 0.2 Jy under optimal conditions. Data
points with a signal-to-noise ratio $<$4 are treated as non-detections.
The flux density scale is set by observations of thermal radio source DR 21. Sources NGC 7027,
3C 274 and 3C 84 are used as secondary calibrators. A detailed description
of the data reduction and analysis is given in \citet{terasranta98}.
The error estimate in the flux density includes the contribution from the
measurement rms and the uncertainty of the absolute calibration.

\subsection{OVRO Data}

\begin{figure*}
\centering
\includegraphics[width=15.0cm]{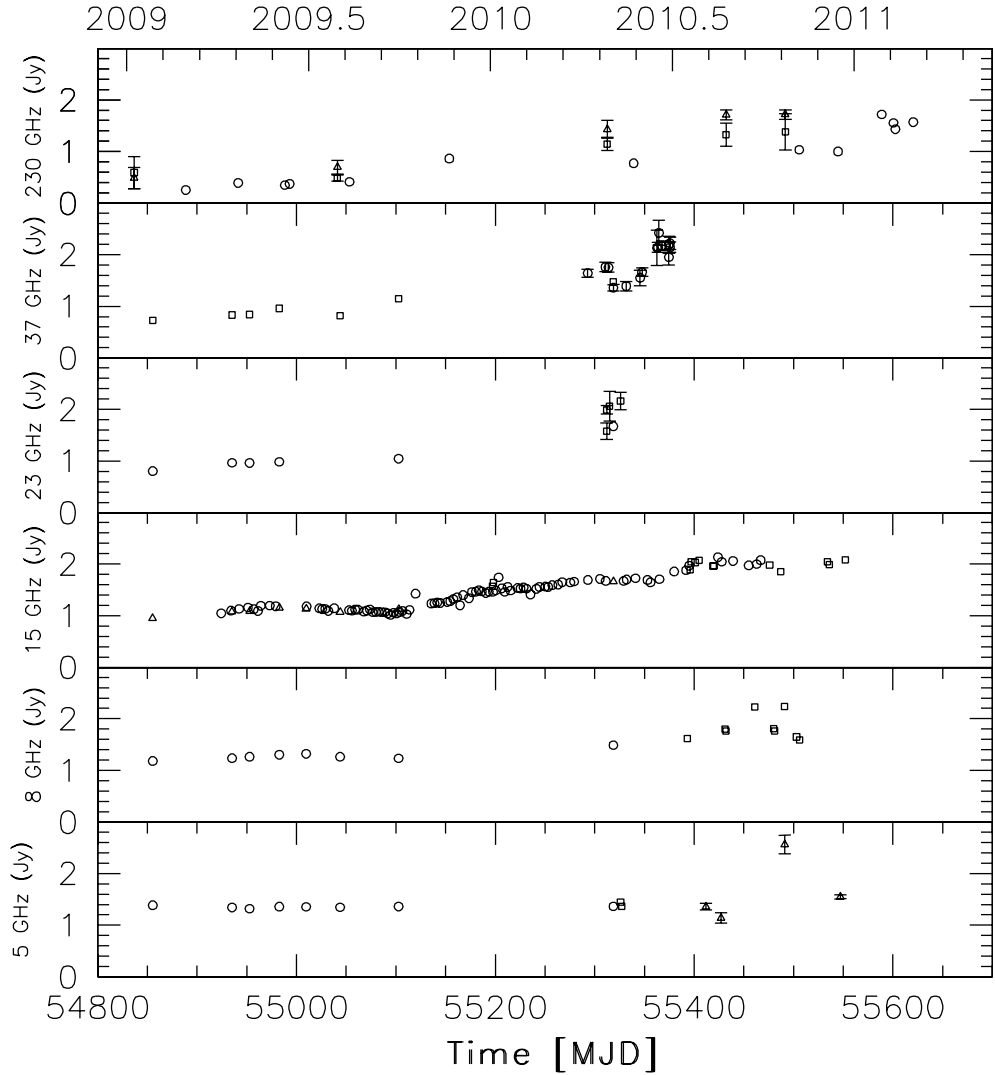}
\caption{Radio and mm light curves of 4C $+$21.35 in units of Jy. The period
  covered is between 2009 January 1 (MJD 54837) and 2011 February 28 (MJD
  55620). The data are collected (from top to bottom panel) by SMA at 230 GHz
  (circles), IRAM at 142 GHz (squares) and 86 GHz (triangles); Mets\"ahovi at
  37 GHz (circles) and Effelsberg at 32 GHz (squares); Effelsberg at 23 GHz
  (circles) and Medicina a 22 GHz (squares); OVRO (circles), UMRAO (squares),
  Effelsberg (triangles) at 15 GHz; Effelsberg (circles) and UMRAO (squares) at 8 GHz;
Effelsberg (circles), Medicina (squares), and UMRAO (triangles) at 5 GHz. For
clarity errors are not shown when $<$5\%.}
\label{radio}
\end{figure*}

As part of an ongoing blazar monitoring program, the Owens Valley Radio
Observatory (OVRO) 40-m radio telescope has observed 4C $+$21.35
at 15~GHz regularly since the end of 2007~\citep{richards11}. This
monitoring program includes about 1700 known or likely $\gamma$-ray-loud blazars,
including all candidate $\gamma$-ray blazar survey (CGRaBS) sources above declination $-20^{\circ}$.  The sources in
this program are observed in total intensity twice per week with a 4~mJy
(minimum) and 3\% (typical) uncertainty on the flux density.  Observations are performed
with a dual-beam (each 2.5~arcmin full-width half-maximum) Dicke-switched system using
cold sky in the off-source beam as the reference. Additionally, the
source is switched between beams to reduce atmospheric variations. The
absolute flux density scale is calibrated using observations of
3C~286, adopting the flux density (3.44~Jy) from \citet{baars77}. This results in about a 5\% absolute
scale uncertainty, which is not reflected in the plotted errors. 4C $+$21.35
was variable at 15 GHz during the
OVRO monitoring (Figure~\ref{radio}), with a flux density ranging from 1.01 Jy (at MJD 55094) to 2.13 Jy (at MJD 55423).

\subsection{UMRAO Data}

UMRAO centimeter band total flux density observations were obtained with the University of Michigan 26-m paraboloid located in Dexter, Michigan, USA.
The instrument is equipped with transistor-based radiometers operating at frequencies centered at 4.8, 8.0, and 14.5 GHz with bandwidths of 0.68, 0.79,
and 1.68 GHz, respectively. Dual horn feed systems are used at 8 and 14.5 GHz,
while at 4.8 GHz a single-horn, mode-switching receiver is employed. Each
observation consisted of a series of 8--16 individual measurements over
approximately a 25--45 minute time period, utilizing an on--off observing technique at 4.8 GHz, and an on--on technique (switching the
target source between the two feed horns, which are closely spaced on the sky)
at 8.0 and 14.5 GHz. As part of the observing procedure, drift scans were made
across strong sources to verify the telescope pointing correction curves, and
observations of nearby calibrators (3C 274, 3C 286, and 3C 218) were obtained every 1--2 hr to correct
for temporal changes in the antenna aperture efficiency. 

\begin{figure}[th!]
\centering
\includegraphics[width=13.0cm]{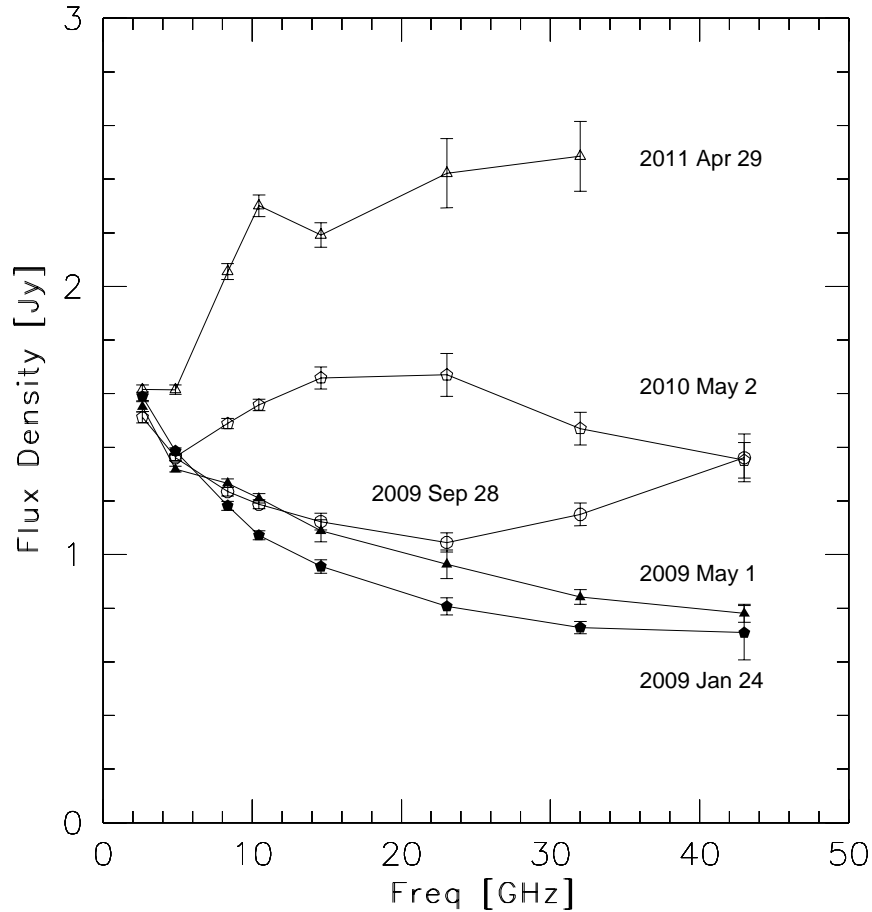}
\caption{Radio spectra of 4C $+$21.35 obtained by Effelsberg on 2009 January
  24 (filled pentagons), 2009 May 1 (filled triangles), 2009 September 28 (open
  circles), 2010 May 2 (open pentagons), and 2011 April 29 (open triangles) from 2.64 to 43 GHz.}
\label{Eff}
\end{figure}

\subsection{Medicina Data}

We observed 4C $+$21.35 with the Medicina radio telescope five times between
2010 April 26 and 2010 May 11. We used the new Enhanced Single-dish Control System (ESCS) acquisition system,
which provides enhanced sensitivity and supports observations with the cross
scan technique. We observed at 22 GHz in the first four epochs, and at 5 GHz in
the last two; the observations on 2010 May 10 were carried out at both
frequencies and can be used for an estimate of the simultaneous spectral
index.

At each epoch, the source was observed for about 10 minutes and calibrated
with respect to  3C 286, after correcting the data for atmospheric opacity. The observing
conditions varied from epoch to epoch, resulting in different noise levels
and significance of the detections. However, after flagging bad scans, we
always obtained a highly significant ($\gg 5\sigma$) detection. The
relative uncertainty on the estimated flux density at 22 GHz varies between
4\% and 15\%, while at 5 GHz it is around 3\%. 

\begin{figure}[!b]
\centering
\includegraphics[width=10.0cm, angle=-90]{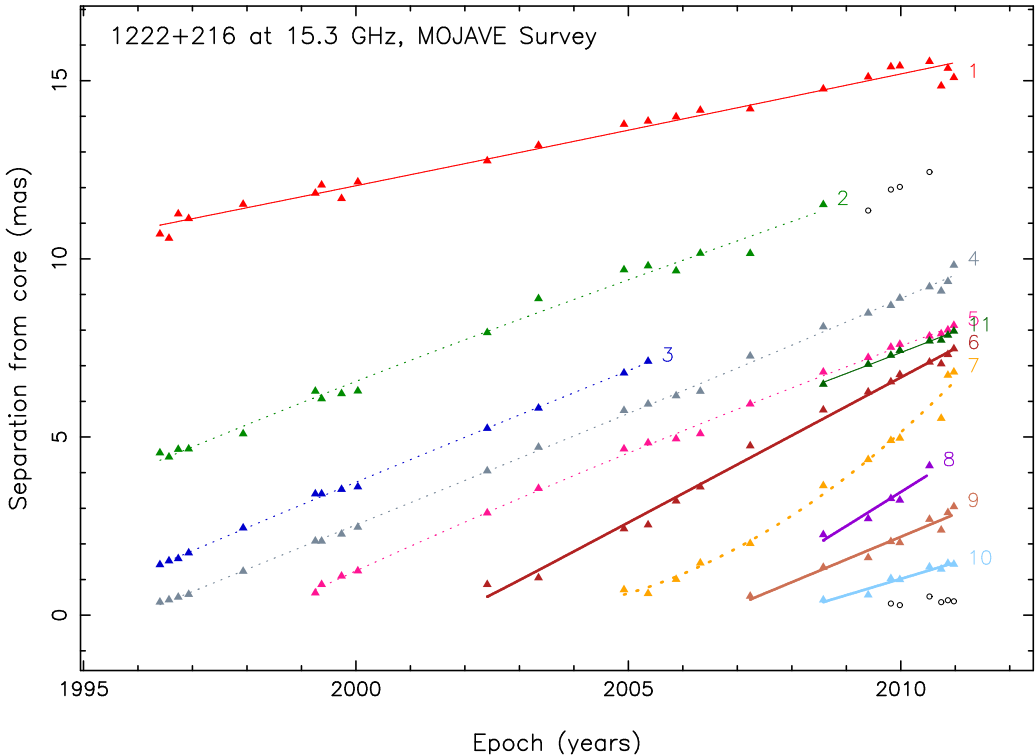}
\caption{Plot of angular separation from core vs. epoch for fitted Gaussian
  jet components in 4C $+$21.35. Color symbols indicate robust components for which kinematic fits were obtained (dotted and solid lines). The solid lines indicate vector motion fits to the data points assuming no acceleration, while the dotted lines indicate accelerated motion fits. Thick lines are used for components whose fitted motion is along a radial direction from the core, while the thin lines indicate non-radial motions. Unfilled black circles indicate non-robust components. The component identification numbers are located next to the last epoch of each robust component.}\label{mojave1}
\end{figure}

\subsection{MOJAVE Data}\label{speeddispersion}

\begin{figure*}[!t]
\centering
\includegraphics[width=17.0cm]{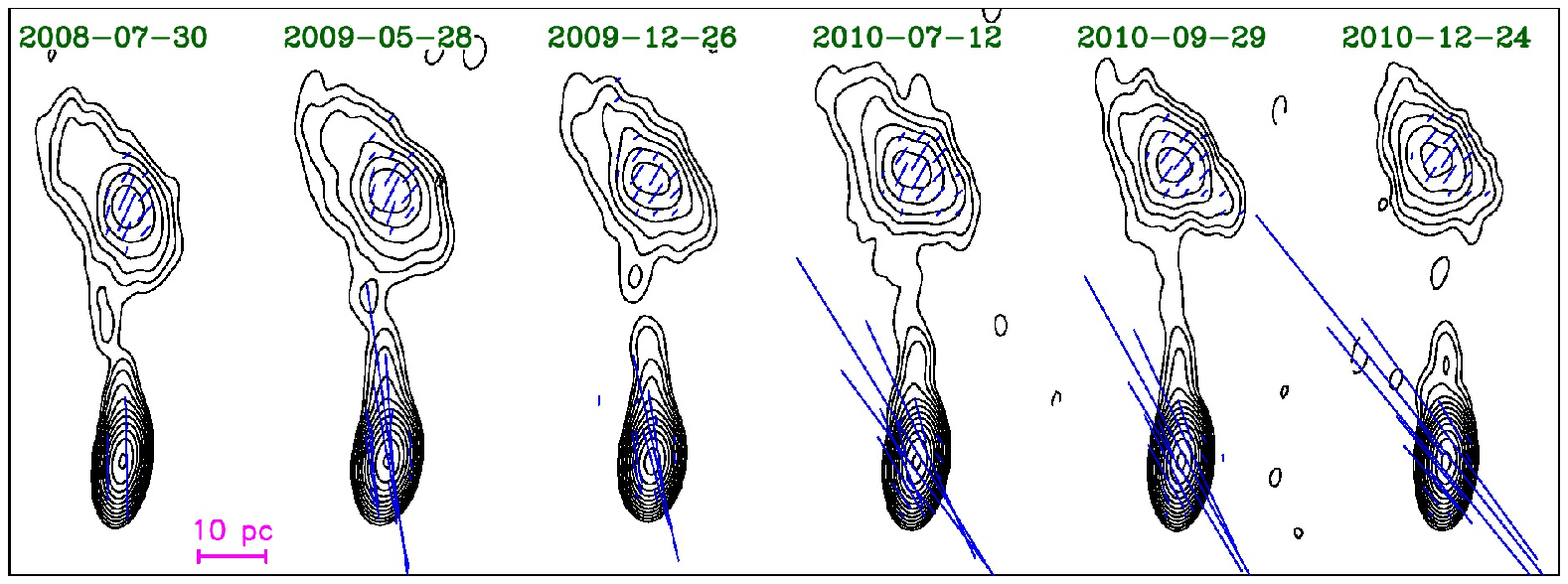}
\caption{Total intensity and linear polarization images of 4C $+$21.35
  observed by VLBA at 15 GHz in different epochs as part of the MOJAVE
  program. Naturally-weighted total intensity images are shown by black contours, the contours are in successive powers of two times the base contour level of 0.2 mJy beam$^{-1}$. Electric polarization vectors direction is indicated by blue sticks, their length is proportional to the polarized intensity.}\label{mojave2}
\end{figure*}

4C $+$21.35 is part of the Monitoring of Jets in Active Galactic Nuclei
with VLBA Experiments (MOJAVE) sample, allowing us to investigate at 15 GHz the radio
structure and jet kinematics of this source over a long period. The data were
processed using the standard procedures described in the AIPS
cookbook\footnote{http://www.aips.nrao.edu} (for details see \citet{lister09}). The radio properties of 4C $+$21.35 strongly indicate that it has a
relativistic jet beamed very closely along our line of sight. The kiloparsec
scale radio morphology from Very Large Array images \citep{cooper07, saikia93}
consists of a bright jet starting out to the north-east of a bright core, and gradually
curving to the east, terminating in a hotspot located approximately 60 kpc from the core.
Surrounding the core is a circular halo of diffuse radio emission $\sim$100 kpc in
diameter, which is consistent with a large radio lobe being viewed end-on. On 
parsec scales, 4C $+$21.35 displays a compact radio jet at an initial position angle of
$\sim 0$$^{\circ}$ that curves roughly 7$^{\circ}$ to the east over 10 mas. However, there
is also a more distant feature at position angle $-6$$^{\circ}$ from the optically thick core.
Multi-epoch Very Long Baseline Array (VLBA) observations by the MOJAVE survey, using data from 1996 until
2011 May \citep[for details of the fitting method see][]{lister13} show that
this outermost feature (id = 1) has an apparent superluminal motion of 8.4$c$,
and is moving to the east (Figure~\ref{mojave1}). Several other jet features closer in have faster
speeds, all close to 17$c$, and are also accelerating to the east. There are two
components (ids=6,7) with even faster speeds of 20$c$ and 27$c$ (Table
\ref{mojave}), that have trajectories curving to the west. These kinematic observations suggest complex three dimensional
trajectories, perhaps having a helical form, which are being investigated in further
detail by the MOJAVE collaboration. The linear fractional polarization and electric vector direction of the core
feature changed between 2009 December and 2010 July, but remained relatively constant from 2010 July to
December (Figure~\ref{mojave2}). There is evidence for a new bright jet feature
in the core region as of 2009 November. The electric vector
directions of the moving features further down the jet were remarkably uniform
with time, pointing in a direction roughly perpendicular to the motion vector
of the outermost moving features. On the other hand, there was no evidence at
15 GHz of a bright superluminal knot ejection during the 2010 $\gamma$-ray flaring period.

\begin{deluxetable}{lcrrrrrrrrrrrrr}
\rotate
\tablecolumns{14}
\tabletypesize{\scriptsize}
\tablewidth{0pt}
\tablecaption{\label{velocitytable}Kinematic Fit Properties of Jet Components}
\tablehead{\colhead {} &   \colhead {} &
\colhead{$\langle S\rangle$}  &\colhead{$\langle R\rangle$} & 
\colhead{$\langle\vartheta\rangle$
} &
  \colhead{$\phi$}&   \colhead{$ |\langle\vartheta\rangle - \phi|$}  
&\colhead{$\mu$}  & \colhead{$\beta_{\rm app}$} &\colhead{$ \dot{\mu}_{\perp} $}  & 
\colhead{$ \dot{\mu}_{\parallel} $}  & &&\colhead{$\Delta \alpha$} &  
\colhead{$\Delta \delta$}  \\
  \colhead {I.D.} &  \colhead {$N$} &
\colhead{(mJy)} &\colhead{(mas)} & \colhead{(deg)}   &
\colhead{(deg)}& \colhead{(deg)} &\colhead{($\mu$as yr$^{-1})$}& \colhead{($c$)}  
&\colhead{($\mu$as yr$^{-2})$} &\colhead{($\mu$as yr$^{-2})$} &\colhead{$T_{ej}$}  
& \colhead{$T_{\rm mid}$} & \colhead{($\mu$as)}& \colhead{($\mu$as)}  \\
\colhead{(1)} & \colhead{(2)} & \colhead{(3)} & \colhead{(4)} &
\colhead{(5)} & \colhead{(6)} & \colhead{(7)} & \colhead{(8)} &
  \colhead{(9)}& \colhead{(10)}&
\colhead{(11)} & \colhead{(12)} & \colhead{(13)} & \colhead{(14)}   & 
\colhead{(15)} }
\startdata

  1 & 24  & 14 &13.5&   $ 354.1$ &  11.5$\pm$1.4 & 17.4$\pm$1.4\tablenotemark{b} 
& 320.4$\pm$9.3 & 8.41$\pm$0.24 &0.049$\pm$0.017 &$-$0.038$\pm$0.021&\nodata 
&2003.84 & 199 &245 \\
    2 & 17  & 20 &7.4&   $ 7.4$ & 15.7$\pm$1.2 & 8.2$\pm$1.2\tablenotemark{b} & 
581$\pm$17\tablenotemark{a} & 15.26$\pm$0.43 
&0.092$\pm$0.019&$-$0.007$\pm$0.026& \nodata &2002.49 & 179 &269 \\
    3 & 13  & 74 &3.7&  $ 2.7$ &  10.19$\pm$0.27 & 
7.48$\pm$0.30\tablenotemark{b} & 640.6$\pm$8.4\tablenotemark{a} & 16.82$\pm$0.22 
&0.0730$\pm$0.0055&0.005$\pm$0.015& \nodata &2000.88 & 29 &93 \\
    4 & 25  & 101 &5.3& $ 4.3$ &  9.71$\pm$0.39 & 5.41$\pm$0.46\tablenotemark{b} 
& 638.9$\pm$5.1\tablenotemark{a} & 16.77$\pm$0.13 
&0.0576$\pm$0.0053&0.0094$\pm$0.0063& \nodata &2003.84 & 114 &136 \\
    5 & 20  & 71 &5.2&   $ 359.5$ & 5.73$\pm$0.21 & 
6.22$\pm$0.23\tablenotemark{b} & 631.2$\pm$5.8\tablenotemark{a} & 16.57$\pm$0.15 
&0.0743$\pm$0.0034&$-$0.0225$\pm$0.0082& \nodata &2005.26 & 43 &109 \\
    6 & 16  & 24 &5.1&   $ 355.8$ & 353.19$\pm$0.99 & 2.6$\pm$1.1 & 779$\pm$15 & 
20.45$\pm$0.40 &$-$0.040$\pm$0.020&0.012$\pm$0.025& 2001.53$\pm$0.14 &2006.84 & 
153 &171 \\
    7 & 14  & 19 &4.1&   $ 4.5$ &  2.96$\pm$0.94 & 1.5$\pm$1.1 & 
1013$\pm$39\tablenotemark{a} & 26.6$\pm$1.0 &$-$0.105$\pm$0.033&0.312$\pm$0.078& 
\nodata &2008.10 & 136 &323 \\
    9 & 10  & 8 &2.1&   $ 0.9$ &  0.16$\pm$0.87 & 0.71$\pm$0.92 & 632$\pm$45 & 
16.6$\pm$1.2 &0.012$\pm$0.038&0.23$\pm$0.15& 2006.55$\pm$0.25 &2009.26 & 36 &168 \\
    10 & 9  & 26 &1.1& $ 4.3$ &  6.9$\pm$2.6 & 2.6$\pm$2.7 & 444$\pm$35 & 
11.67$\pm$0.92 &\nodata&\nodata& 2007.82$\pm$0.20 &2009.93 & 49 &87 \\
    11 & 9  & 27 &7.5& $ 357.8$ &  17.7$\pm$2.3 & 19.9$\pm$2.3\tablenotemark{b} 
& 615$\pm$23 & 16.15$\pm$0.60 &\nodata&\nodata& \nodata &2009.93 & 61 &56 \\
    12 & 7  & 99 &0.4& $ 8.4$ &  18.7$\pm$4.5 & 10.3$\pm$4.5 & 96$\pm$19 & 
2.52$\pm$0.49 &\nodata&\nodata& \nodata &2010.55 & 6 &26 \\
\enddata
\tablenotetext{a}{Component shows significant accelerated motion.}
\tablenotetext{b}{Component shows significant non-radial motion.}

\tablecomments{The kinematic fit values are derived from the acceleration fit
  for components with significant acceleration, and from the vector motion fit
  otherwise. Columns are as follows: (1) component number, (2)
  number of fitted epochs, (3) mean flux density at 15 GHz in mJy,  (4) mean
  distance from core component in mas, (5) mean position angle with respect to the core component
  in degrees, (6) position angle of velocity vector in degrees, (7) offset
  between mean position angle and velocity vector position angle in degrees,
  (8) angular proper motion in microarcseconds per year, (9) fitted speed in units of the speed of light,
  (10) angular acceleration perpendicular to velocity direction in
  microarcseconds per year per year, (11) angular acceleration parallel to velocity
  direction in microarcseconds per year per year, (12) fitted ejection date, (13) date of reference (middle) epoch used for fit, (14) right ascension error of individual epoch positions  in $\mu$as, (15) declination error of individual epoch positions in $\mu$as.}\label{mojave}
\end{deluxetable}

\section{Discussion}

\begin{figure*}[p]
\centering
\includegraphics[width=15cm]{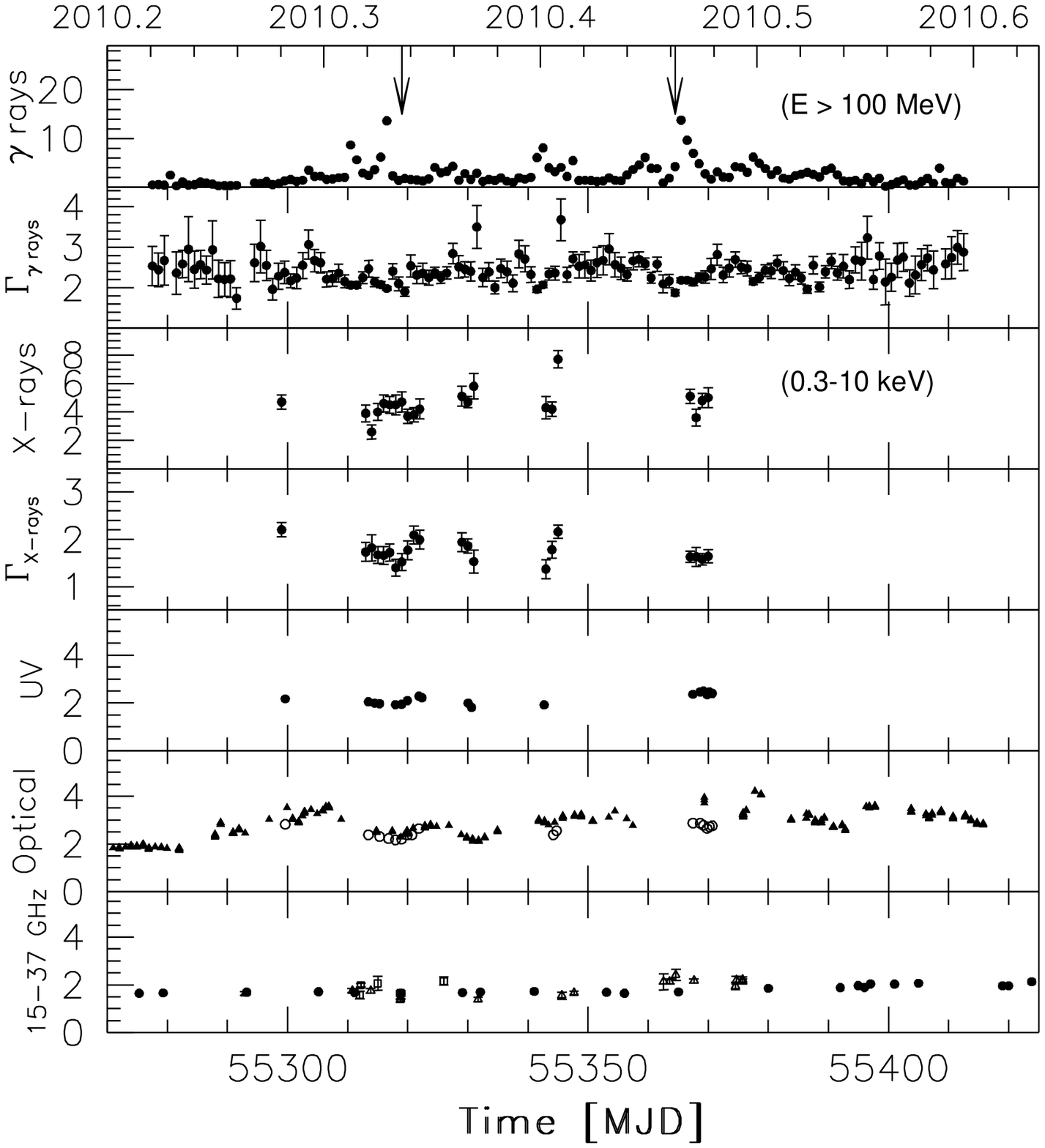}
\caption{Multifrequency light curves of 4C $+$21.35 between 2010 April 9 and August
4 (MJD 55295--55412). The data sets were collected (from top to
bottom) by {\em Fermi}-LAT \citep[$E>$ 100 MeV; in units of 10$^{-6}$ photons cm$^{-2}$ s$^{-1}$; taken from]{tanaka11}, {\em Swift}-XRT (0.3--10 keV; in units of 10$^{-12}$ erg
cm$^{-2}$ s$^{-1}$), {\em Swift}-UVOT (m2 filter; in units of mJy), {\em Swift}-UVOT
(U filter, open circles; in units of mJy), Abastumani, ATOM, Crimean, KVA, St.~Petersburg
($R$-band, filled triangles; in units of mJy), Effelsberg, Medicina, Mets\"ahovi, OVRO, UMRAO (15 GHz: filled
circles, 23 GHz: open squares, 37 GHz: open triangles; in units of Jy). The downward
arrows indicate the times of the VHE detections by MAGIC. For clarity the
$m2$, $u$, $R$ and 15 GHz bands errors (typically 5\% or less) and the $\gamma$-rays errors are not shown.}
\label{MWL}
\end{figure*}

\subsection{Light Curves Behavior and Correlation}\label{lc}

\begin{figure}[!th]
\centering
\includegraphics[width=12.0cm]{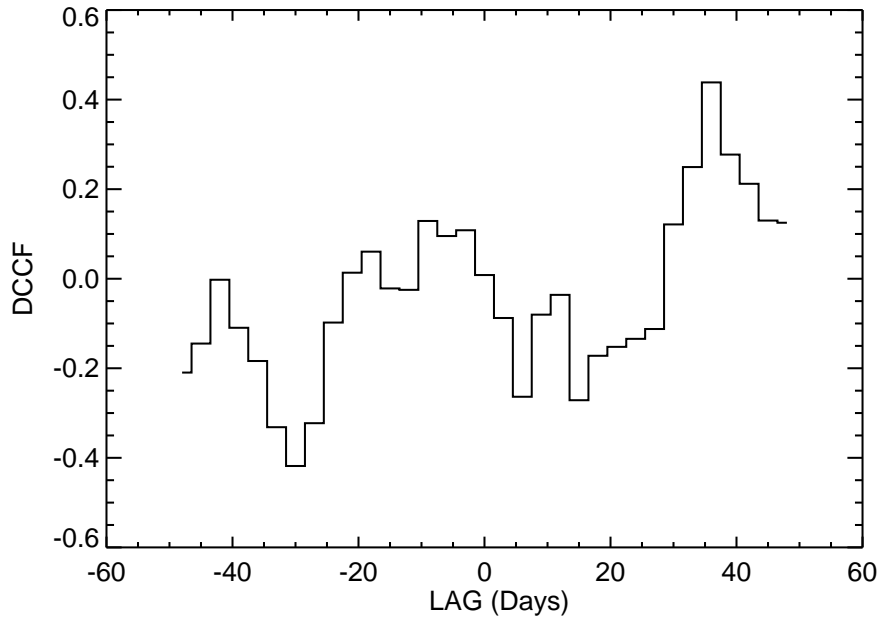}
\caption{Discrete cross correlation function between the $\gamma$-ray and
  $R$-band light curves of 4C $+$21.35.}
\label{DCCF}
\end{figure}

\begin{figure}[!th]
\centering
\includegraphics[width=15.0cm]{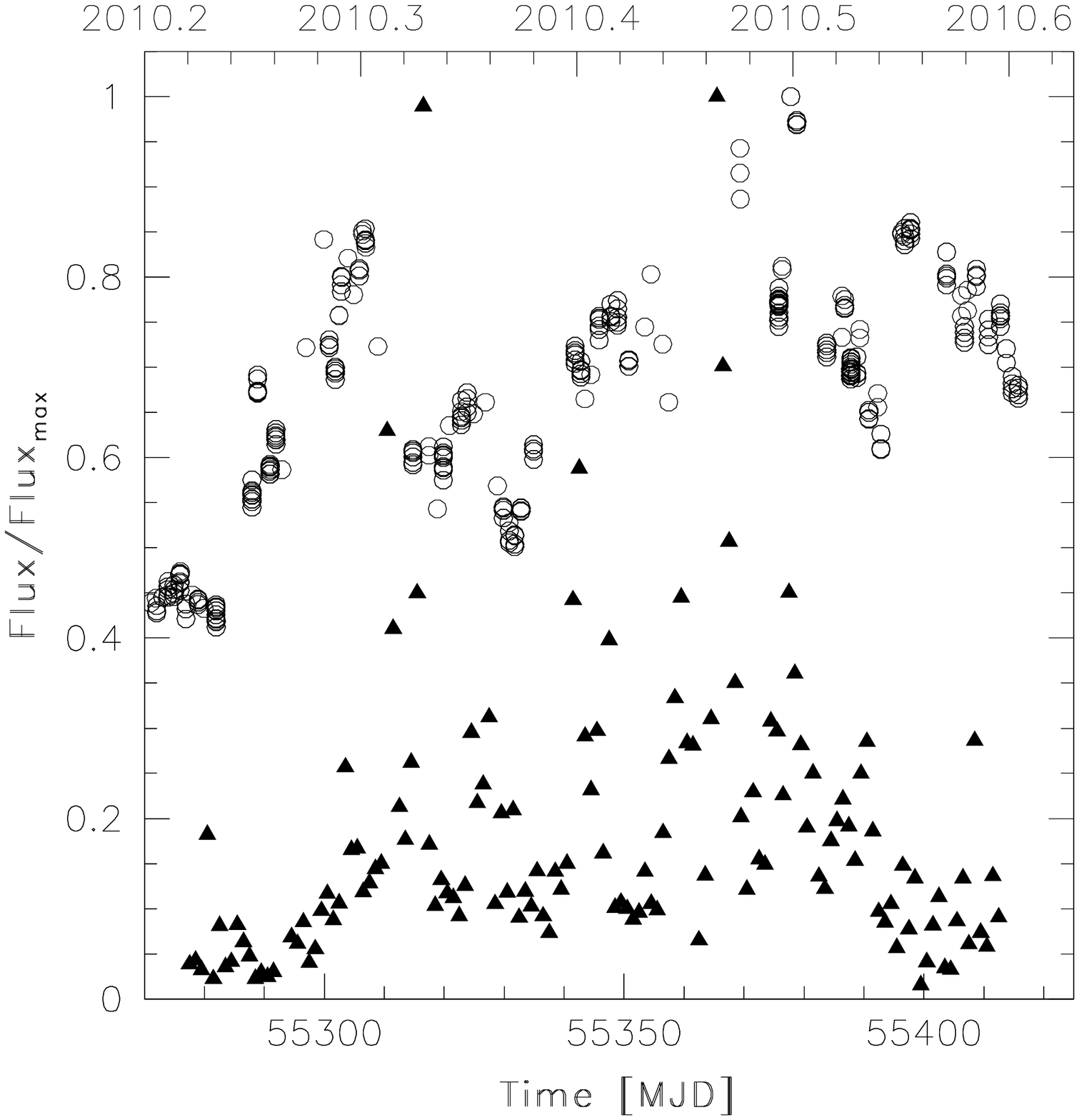}
\caption{Comparison between $\gamma$-ray and $R$-band light curves. We superimpose $\gamma$-ray
(black triangles) and $R$-band (red empty circles) light curves normalizing $\gamma$-ray and $R$
flux values over the whole observing period to the respective peak flux values.}
\label{comparison}
\end{figure}

The multifrequency light curve of 4C $+$21.35 in Figure~\ref{MWL} shows
the {\em Fermi}-LAT, {\em Swift} (XRT; UVOT, $u$ and $m2$ filters), optical $R$-band
(Abastumani, ATOM, Crimean, KVA, St.~Petersburg), and radio from 15 GHz to 37
GHz (Effelsberg, Medicina, Mets\"ahovi, OVRO, UMRAO) data collected during 2010 April 9--August 4 (MJD
55295--55412). In addition the $\gamma$-ray and X-ray photon indices observed by {\em Fermi}-LAT and {\em Swift}-XRT are reported in the second and fourth panels.
The {\em Fermi}-LAT light curve shows two distinct
$\gamma$-ray flaring episodes, peaking on 2010 April 29 (MJD 55315)
and June 18 (MJD 55365), together with other peaks of lesser
brightness. The two major $\gamma$-ray peaks detected by {\em
Fermi}-LAT occurred very close in time to the VHE detections by
MAGIC, on May 3 and June 17. This indicates that the same emission
mechanism may be responsible for both the HE and VHE emission during
these flaring episodes, in agreement also with the fact that the
combined HE and VHE spectrum in 2010 June 17, corrected for the EBL
absorption, can be described by a single power-law \citep{MAGIC_discovery}. It is also worth noting
that during the two VHE detections the photon index estimated in the LAT
energy range is quite flat ($\Gamma_{\gamma}$ $\sim$2), favoring the
detection of $\gamma$-ray emission up to hundreds of GeV.

During 2010 February--June, {\em Swift}/XRT observed 4C $+$21.35 with
a 0.3--10 keV flux in the range (2.6--7.7)$\times$10$^{-12}$ erg
cm$^{-2}$ s$^{-1}$, with the photon index changing in the range
1.4--2.2. The photon index remained constant during the 2010 April and June $\gamma$-ray flaring periods (see Figure~\ref{MWL}).
The very small variability amplitude observed in X-rays
($\sim$3) with respect to the MeV--GeV energy range ($\sim$70) could indicate that the low-energy segment of the electron energy
distribution responsible for the production of the X-ray photons
varies much less than the high-energy electron tail
involved in the production of the observed $\gamma$-ray emission. A
small variability amplitude was observed in UV during 2010. This could
be due to the fact that the UV part of the spectrum is dominated by
the accretion disk emission that dilutes the jet emission. It is worth
noting that a peak of the UV emission was detected on June 18, but the
small increase observed makes it unlikely that the change of the
accretion rate is the main driver of the simultaneous activity
observed at the higher energies by MAGIC and {\em Fermi}-LAT.

The $R$-band light curve is quite well sampled and shows variable flux
density over time, but no dramatic increase of the activity. Two
optical peaks were observed on 2010 April 20 (MJD 55306) and 2010 June
30 (MJD 55377), close in time but not simultaneous with the two
$\gamma$-ray peaks. For the second flaring event, the lack of strictly simultaneous ground-based optical observations was
covered by the UVOT observations that seems to indicate a relatively
high activity at MJD 55367 (June 20).  Correlations between the
$\gamma$-ray and optical light curves of 4C $+$21.35 were investigated
by computing the discrete cross correlation function (DCCF), following
\citet{edelson88} and \citet{white94} (see Figure~\ref{DCCF}; positive lag means
that $\gamma$-ray flux variations occur before those in $R$-band; the DCCF value ranges from --1 to +1). Although
the overall $R$-band flux was higher during the period of $\gamma$-ray
activity (see Figure~\ref{comparison}) the DCCF shows no clear evidence
for correlations on the timescale of the rapid flares
($\sim$days), with a maximum correlation of 0.4 for a time-lag of $\sim$35 days. A similar conclusion was reached by \citet{smith11} from a comparison of
a LAT light curve during this epoch with the Steward Observatory
$V$-band observations also used in this paper. In particular, overall correlation between the $\gamma$-ray band with the $R$-band was
observed for 4C 21$+$35 during the 2010 $\gamma$-ray flaring activity, but on short time
scales some differences are evident (see Figure~\ref{comparison}). 
A complex connection between
the optical and $\gamma$-ray emission has been already observed in several
FSRQs and low-synchrotron-peaked BL Lac objects. In some cases a clear optical/$\gamma$ correlation with no lags was observed \citep[e.g., 3C
279;][]{abdo10_3c279}. But sometimes no correlation was found between
  these two energy bands \citep[e.g., BL Lacertae;][]{abdo11}, and in other
  occasions, an optical and NIR flare with no significant counterpart in
  $\gamma$-rays was observed \citep[e.g., PKS\,0208$-$512 and PKS\,0537$-$441;][]{chatterjee13,dammando13_0537}. 
 
An increasing flux density was observed in radio and mm bands from the
beginning of 2009 (see Figure~\ref{radio}) contemporaneous with the
increasing $\gamma$-ray activity observed by {\em Fermi}-LAT, reaching the peak of flux density at 230 GHz on 2011 January 27 (MJD
55588). Interestingly, the peak of the 23 GHz and 37 GHz was
observed on 2010 May 10 (MJD 55326) and June 18 (MJD 55365),
respectively, close in time with the major $\gamma$-ray flares. The
same activity was also observed at 8 GHz and 5 GHz, with the emission
peak delayed likely due to synchrotron self-absorption opacity effects. However, the sparse coverage does
not allow us to obtain conclusive evidence. A significant spectral evolution was also observed in radio (see Figure \ref{Eff}), with the spectrum
  changed from steep on 2009 January 24 ($\alpha_r$ = 0.3) to inverted ($\alpha_r$ = --0.2) on 2011 April 29 (see Section \ref{effelsberg}).

\subsection{SED Modeling}
\label{section:SEDmodel}

\subsubsection{Data Selection}

\begin{figure*}
\centering
\includegraphics[width=15.0cm]{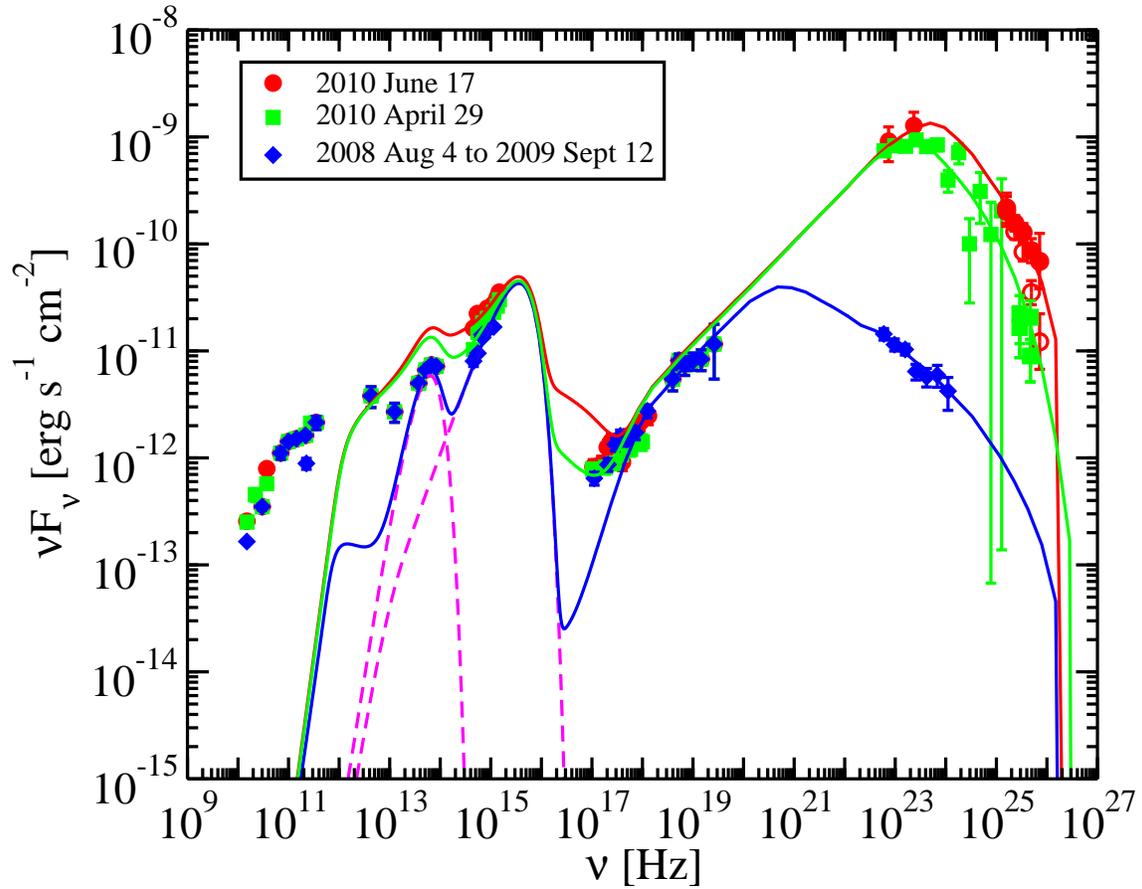}
\caption{Spectral energy distribution of 4C $+$21.35 in three epochs:
  2010 June 17 (red circles), 2010 April 29 (green squares), and 2008
  August 4--2009 September 12 (blue diamonds). Dashed magenta lines
  indicate the dust torus and accretion disk emission
  components. The MAGIC data have been corrected for EBL absorption using the model of \citet{finke10}. Empty symbols refer to non-EBL-corrected data, filled symbols to EBL-corrected ones.}
\label{SED}
\end{figure*}

We have built three quasi-simultaneous SEDs from the data
discussed above, shown in Figure~\ref{SED}. These SEDs include the
flaring states of 2010 June 17 (red circles) and 2010 April 29 (green
squares), and a quiescent state, integrated in time from 2008
August 4 to 2008 September 12 (blue diamonds).
For the three SEDs we used the LAT spectra calculated over 2010
June 17, 2010 April 23--May 2, 2008 August 4--2009 September 12
\citep[taken from][]{tanaka11,MAGIC_discovery}, and the {\em Swift}
data collected on 2010 June 20, 2010 April 23, and 2009 April 10,
respectively. The MAGIC data collected on 2010 May 3 and June 17 are included for the two flaring states. These data have been
corrected for EBL absorption using the model of \citet{finke10}. This
model is nearly identical in the energy range covered by MAGIC to the
model of \citet{dominguez11} used in \citet{MAGIC_discovery}.  We also
included the radio and $R$-band observations nearest to the LAT
$\gamma$-ray peak for the two flaring states (2010 April 28 and June
22, respectively), and the {\em Swift} observation performed on 2009
April 19 for the quiescent state. Finally we report in the SEDs the
average {\em Swift} BAT 70-month spectrum, the {\em Planck} spectrum
collected in 2009 December \citep{planck}, and the {\em Spitzer} data
from \citet{malmrose11}.

\noindent \citet{farina12} estimate the black hole (BH) mass for 4C~+21.35 as
$M_{\rm BH}$ $\sim$ 6$\times10^8\ M_\odot$, using broad emission line measurements
from over 100 optical spectra from a variety of sources.  This value agrees with values found by \citet{shen11} and
\citet{shaw12} with an Sloan Digital Sky Survey (SDSS) spectrum, but significantly greater than
the value found by \citet{wang04} and used by \citet{tanaka11},
$M_{\rm BH}$ $\sim$ 1.5 $\times 10^8 M_{\odot}$. It is worth noting that the measurement by \citet{wang04} relies on the H-$\beta$ broad
line and the continuum luminosity-BLR radius relation from \citet{kaspi00}. That relation was obtained from reverberation mapping of a
small number of active galactic nuclei using a cosmology with a decelerating universe, which is now known to be incorrect.  
We use the more precise value from \citet{farina12} in our SED modeling.

\subsubsection{Model}

We model the SED of the three epochs using a one-zone leptonic model. 
We began our modeling efforts by attempting to model the IR data from
\citet{malmrose11} with a blackbody dust torus.  The results for the
luminosity ($L_{\rm dust}$) and temperature ($T_{\rm dust}$) of the blackbody
were similar to the ones found by \citet{malmrose11}.  The optical
emission clearly appears to be dominated by thermal disk emission,
rather than nonthermal synchrotron emission from the jet, otherwise
the optical spectrum would appear much softer.  Therefore, we next
modeled the optical data in the low-state with a Shakura--Sunyaev
multi-temperature disk \citep{shakura73}, assuming $M_{\rm BH}$ $\sim$
6$\times10^8\ M_\odot$.  We note that the disk fit to the low-state
data is insensitive to the inner radius of the disk, $R_{in}$, as seen
in a close up of this part of the SED in Figure~\ref{SEDoptical}.
Parameters for the dust torus, accretion disk, and all other modeling
parameters can be found in Table \ref{table_fit}. \cite{tav11} use an
  isotropically emitting blackbody spectrum to fit the blue bump and obtain a value for
  the disk luminosity over twice the value presented here. We use a
Shakura--Sunyaev disk, which does not emit isotropically, and which we assume
emits as the cosine of the disk inclination angle. With this distribution, for a face-on disk, the flux will be
twice that from an isotropic distribution for a given luminosity
\citep[e.g.,][]{castignani13}. This is the cause of most of the discrepancy, with the remaining discrepancy due to the different contributions
from nonthermal synchrotron emission.

Although several possibilities have been suggested for the origin
of $\gamma$-ray emission from 4C+21.35 (see Section \ref{intro}),
FSRQ-type blazars such as 4C~+21.35 are expected to have their $\gamma$-rays originate from the external Compton (EC), rather than
synchrotron self-Compton (SSC) mechanism
\citep[e.g.,][]{ghisellini98}. Therefore we next attempt to fit the SED
in the high state of 2010 June 17 (MJD 55364) with a combination of
synchrotron, SSC, and EC emission from a jet blob moving at a highly
relativistic speed. We assume an emitting size of $R^{\prime}_b=10^{15}$\ cm in the comoving frame, consistent with the rapid variability timescale of 10 minutes. The dust torus and disk emission are not varied
between flaring and quiescent states.  For the nonthermal jet emission
we choose a variability timescale of 10 minutes, consistent with the
variability observed by MAGIC \citep{MAGIC_discovery}.  The MAGIC
detection of the source out to $\ga 300$\ GeV also implies the primary
emitting region must be outside the BLR, otherwise $\gamma\gamma$
absorption by broad-line photons would not allow such high-energy
$\gamma$-ray photons to escape \citep{tanaka11,MAGIC_discovery}, so we chose a large jet distance from the BH, $r$, outside the BLR
radius of $R_{\rm BLR}\approx 2\times10^{17}$\ cm \citep{tanaka11}.
Outside the BLR, the seed photon source is expected to be from the
dust torus, which is what we use as the EC seed photon source.  For
the purposes of calculating the geometry of Compton scattering, we
assume the dust torus is a one-dimensional ring with radius
$R_{\rm dust}$, aligned orthogonal to the jet, where we choose $R_{\rm dust}$
to be roughly consistent with the value of the dust sublimation radius
calculated by \citet{nenkova08}.  This is necessary since our
calculations use the full angle-dependent Compton cross section,
accurate in the Thomson through Klein-Nishina (KN) regimes. The adopted synchrotron component is self-absorbed below ∼10$^{12}$ Hz.
We treat the radio points as upper limits, since their slow
variability compared to the optical and $\gamma$-ray emission and flat
spectrum (in flux density $F_{\nu}$) imply they are probably the
result of a superposition of several self-absorbed jet components
\citep{konigl81}, and not the result of the same emitting region that produces the rest of the SED.  The electron distribution was
assumed to be a broken power-law between electron Lorentz factors
$\gamma_{\rm min}$ and $\gamma_{\rm max}$ with power-law index $p_1$ for
$\gamma<\gamma_{\rm brk}$ and $p_2$ for $\gamma>\gamma_{\rm brk}$.  Further
details on the model and its parameters can be found in
\citet{finke08_ssc} and \cite{dermer09}.

The result of this fit to the 2010 June 17 (MJD 55364) SED is
shown in Figure~\ref{SED}. We note that there is some degeneracy in the choice of the model parameters, hence the set of parameter values describing the observational data are not unique. However, we do demonstrate that a one-zone model can adequately describe the data.
 To account for the highest speeds
derived by the jet kinematics analysis of the MOJAVE data (see Section
\ref{speeddispersion}) at least some portion of the jet must be viewed
within $\sim$4$^{\circ}$ of the line of sight. To avoid the
extreme KN regime for Compton scattering, we found that the jet needs to be
highly aligned, with the jet angle with respect to the line of sight
$\theta \approx 0^{\circ}$ ($\delta_D \approx2\Gamma$), where
$\delta_D$ is the Doppler factor. This is because the energy at
which the extreme KN regime begins is at $\epsilon_{\rm KN} \approx
(\delta_D/\Gamma) \epsilon_0$, where $\epsilon_0$ is the seed photon
energy. Such a small jet's angle does not disagree with the high
apparent speeds estimated on the scales of a few parsecs if the
complex three dimensional trajectories observed by MOJAVE are taken
into consideration. In fact, there is evidence for a bend in the jet
on the parsec scale in the VLBA images, where the emission in this model originates (see Section \ref{speeddispersion}).  The
model does not provide a good fit to the XRT data in this SED,
with the model being dominated by synchrotron emission for the soft
X-rays, while the XRT spectral index is $\Gamma_{\rm X}<2$ indicating
it is dominated by some sort of Compton scattering, either SSC or EC
(EC in the case of our model fit).  However, the XRT data were
not strictly simultaneous with the rest of the SED, particularly the
LAT data (with a gap of 6 and 3 days between the X-ray and
$\gamma$-ray data, respectively). As can be seen in Figure~\ref{MWL}, the XRT photon index alternates between $\Gamma_{\rm X} < 2$,
implying Compton scattering dominates in this waveband, and $\Gamma_{\rm X} > 2$, implying
synchrotron dominates. If the primary emitting region
makes up the majority of the jet cross section, this model fit gives a
jet half-opening angle of $\theta_{\rm open} \sim R^{\prime}_b/r \sim
10^{-4}\ \rm{rad} \sim$ 0\fdg007, where $R^{\prime}_b$ is the
comoving radius of the blob.  Such a small opening angle is highly
unlikely and inconsistent with radio observations
(Figure~\ref{mojave2}), so this model implies that the overwhelming
majority of the source's emission is coming from a very small fraction
of the jet's cross section.  We also calculated the jet power in
electrons ($P_{j,e}$) and Poynting flux ($P_{j,B}$) for this model
fit, assuming a two-sided jet \citep{finke08_ssc}, finding that the
source has almost 100 times as much power in electrons as in Poynting
flux.  The model fit to the 2010 June 17 (MJD 55364) flaring SED is
similar to the ``case A'' fit to the same SED data by \citet{tav11}.
They also found a jet where the electron energy density dominates over the magnetic energy density,
although in their case it is even more dominant, with $P_{j,e} \sim
10^4 P_{j,B}$.  \citet{tav11} also provide two other fits to the same
SED with two zone models: a ``case B'' where there is an additional
contribution from a larger blob outside the BLR; and a ``case C''
where there is a contribution from a larger blob inside the BLR.
Neither of these two-zone fits solves the problem of having an
extremely small, bright blob at a large distance from the BH, although
they do provide fits much closer to equipartition between electrons
and Poynting flux.  The UV data for the 2010 June 17 flaring SED
requires an inner disk radius $R_{in}< 6 R_g$, with the
best fit found for $R_{in}=3 R_g$ ($R_g$ is the gravitational radius).
We discuss the implications of this below.

First, however, we discuss the fit to the other bright flare, on 2010
April 29.  This SED is quite similar to the 2010 June 17 one, and we
found we could fit this SED with only minor changes in the
electron distribution, keeping the other parameters the same.
Specifically, this required lowering $\gamma_{\rm brk}$ from
$1\times10^{3}$ to $6\times10^2$ and $\gamma_{\rm max}$ from $4\times10^4$
to $2\times10^4$.  This resulted in a slightly lower $P_{j,e}$, as
seen in Table \ref{table_fit}.  For this flaring state, the lower
$\gamma_{\rm max}$ yields a better fit for the XRT data. The UV data for this
state also are more consistent with an inner disk radius
$R_{in}=3 R_g$. 

Finally, we turn to the ``quiescent state'' SED, derived by integrating
  LAT data from 2008 August 4 to 2009 September 12 in addition to multifrequency
  data in the same period.  We again find a good fit changing only the electron distribution
parameters from the flaring states, while keeping the rest of the
parameters the same.  Here we varied the electron break to
$\gamma_{\rm brk}=26$, and changed the normalization, keeping all other
parameters the same as the fit to the 2010 June 17 flaring state.
This provides a good fit to the SED, although it presents some
peculiarities. In this model, the synchrotron peak would be observed
at frequency $\nu_{pk} \approx m_ec^2/h\ \gamma_{\rm brk}^2 B/B_{\rm crit}
\delta_D/(1+z) \approx 7.4\times10^{10}$\ Hz, if this part of the
spectrum is not highly synchrotron self-absorbed.  Instead the peak
is at $\sim 10^{12}$\ Hz, at the self-absorption frequency, where the
model flux is about an order of magnitude below the data.
This is not strictly a problem, since the observed radio emission is
probably from a much larger region of the jet, but it does seem
strange to have such a low synchrotron peak frequency.  For the fit to the
quiescent state, the model underpredicts the softest XRT flux, rather
than overpredicting it as the model for the 2010 June 17 flare did. Again, this could be due to variability during this rather long
quiescent time period.  It is also possible that the X-ray emission
originates from a different region, maybe even from an accretion disk corona, particularly since the accretion disk is so
prominent. There have been many instances in FSRQs where the X-ray continuum has been characterized by very distinct variability
properties compared to optical and $\gamma$-ray flares \citep[e.g.,][]{abdo10_3c279,marscher10}.

\subsubsection{Accretion Disk Emission}

\begin{figure*}
\centering
\includegraphics[width=15.0cm]{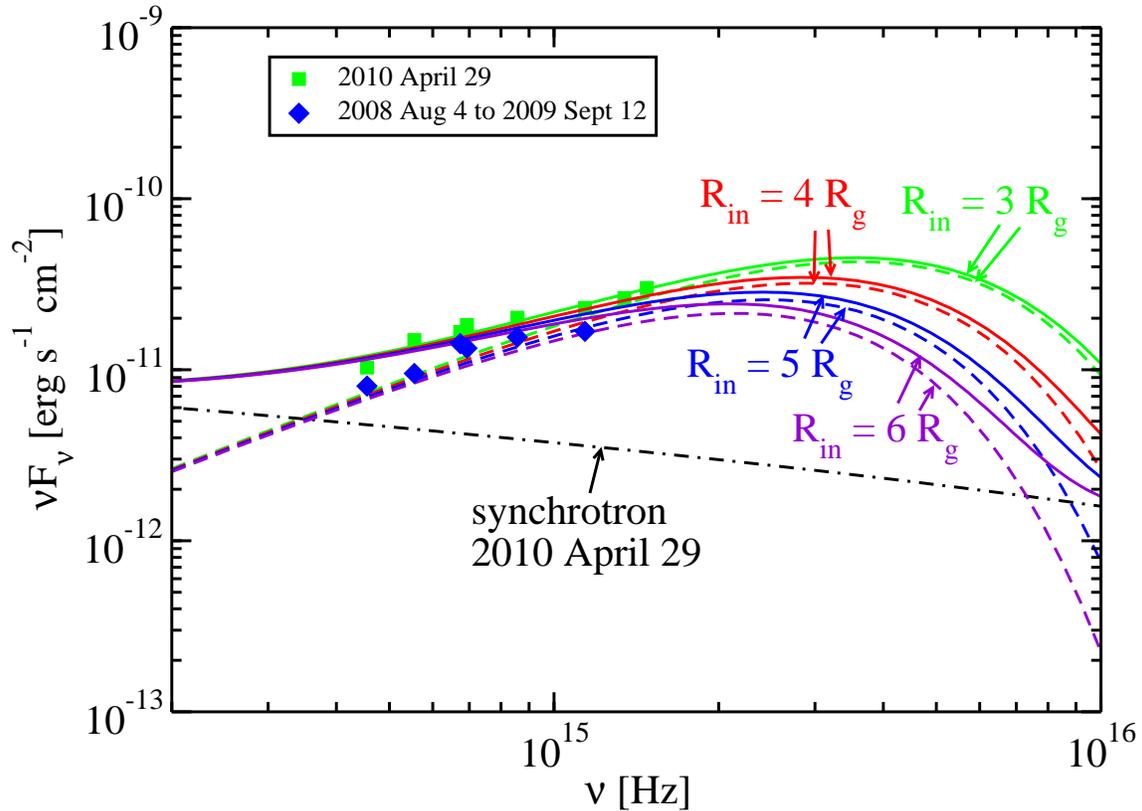}
\caption{Similar to Figure~\ref{SED}, but zoomed in on the optical
portion of the spectrum, which in our model originates mainly from
disk emission.  Model disk emission for several inner disk radii are
shown (dashed curves), while the synchrotron from the model fit of
2010 April 29 is shown as the dot-dashed curve.  The total
(synchrotron + disk) emission is shown as the solid curves.  Models
with large inner disk radii do not provide an adequate fit to the UV
data of 2010 April 29.}\label{SEDoptical}
\end{figure*}

For both the flaring state models, we find that a good fit to
the UV data from {\em Swift}-UVOT requires an inner disk radius
$R_{in}<6 R_g$, where $6 R_g$ is the value one would expect for the
innermost stable circular orbit around a nonrotating Schwarzschild BH.
Indeed, our fits favor $R_{in}=3 R_g$ (see Figure~\ref{SEDoptical}), the value one would expect for a maximally
(prograde) rotating Kerr BH.  This is because, as seen in the
figure, a larger $R_{in}$ will not fit the UV data points. We also
performed fits with the color correction of \citet{chiang02}. This
requires a slightly higher disk luminosity ($L_{\rm disk} =
2\times10^{46}$ erg s$^{-1}$), but our results for the
inner disk radius remain unchanged.  This is of interest since one
would expect a BH spin at or near the maximum value if the jet is
generated from the Blandford--Znajek mechanism \citep{blandford77}. It is also inconsistent with the scenario of \citet{garofalo10},
where the jets in powerful FR II sources (and presumably FSRQs) are
produced by BHs with retrograde spin, requiring that $R_{in}>6 R_g$,
while jets in less powerful FR I sources (and presumably BL Lac objects) are
produced by BHs with prograde spin. However, there are some caveats regarding the fit of the optical/UV data. The results depend
on the BH mass, although the results for this seem to converge  to around
6--8$\times10^8\ M_{\odot}$.  We also note that if the BH mass were as
low as the one found by \citet[][]{wang04}, $M_{\rm BH}=1.5\times10^8\
M_{\odot}$, we would not be able to fit the blue bump for this source
with a value of the disk luminosity $L_{\rm disk}$ that is less than the
Eddington luminosity.  The UV data are often subject to heavy
extinction, which could lead to large uncertainty.  If the synchrotron
component was less steep it could also potentially have a greater
contribution to the UV region, masking a larger $R_{in}$.  But in this
case the synchrotron emission would over-predict the longer wavelength
optical data, so this is unlikely.  Finally, the disk model we use is
rather simple.  It does not include a general relativistic effects
such as gravitational Doppler shifts or light bending \citep{li05}.

Are our modeling results consistent with the observed optical spectra 
of 4C+21.35?
Estimates for the luminosity of the broad H$\beta$ line range from $L_{\rm H\beta} =
2.1\times10^{43}$\ erg s$^{-1}$ \citep{fan06,tanaka11} to the values
found from the SDSS spectrum, $L_{\rm H\beta} =6.3\times10^{43}$\ erg
s$^{-1}$ as measured by \citet{shen11} and $L_{\rm H\beta} =5.5\times10^{43}$\ erg s$^{-1}$ by \citet{shaw12}.  \citet{farina12}
find the line to be quite variable by systematically studying a
variety of spectra at different epochs, and their values range from
$L_{\rm H\beta} =3.7\times10^{43}$\ erg s$^{-1}$ to $L_{\rm H\beta} =6.2\times10^{43}$\ erg s$^{-1}$.  Values for the luminosity at 5100
\AA\ are fairly constant if one is careful to exclude the
nonthermal component.  From the same spectrum, \citet{shen11} find
$L_{5100}=3.8\times10^{45}$\ erg s$^{-1}$ and \citet{shaw12} find
$L_{5100}=3.5\times10^{45}$\ erg s$^{-1}$.  The estimates by
\citet{farina12} varies considerably, but their lowest value, with
presumably the least amount of contribution from the nonthermal
emission, is $L_{5100}=3.5\times10^{45}$\ erg s$^{-1}$.  \citet{greene05}
found a tight correlation between $L_{5100}$ and $L_{\rm H\beta}$, and all the
values here, which are around $L_{\rm H\beta}/L_{5100}\approx
0.01$, are consistent with their correlation.  If the total BLR
luminosity is $L_{\rm BLR}=25.3\times L_{\rm H\beta}$ \citep{tanaka11}, then,
using a value $L_{\rm H\beta}=5\times10^{43}$\ erg s$^{-1}$ that is
consistent with the recent measurements
\citep{shen11,shaw12,farina12} one gets $L_{\rm BLR}=1.3\times10^{45}$\
erg s$^{-1}$. Thus, using the disk luminosity for our model, one gets $\xi_{\rm BLR}\cong L_{\rm BLR}/L_{\rm disk}\approx0.08$, a fairly standard value. Our model fit gives a value of the fraction of the disk radiation reprocessed in the dust torus $\xi_{\rm dust} \cong L_{\rm dust}/L_{\rm disk} = 0.34$, again a fairly standard value \citep[e.g.,][]{sikora09}.

\subsubsection{Jet and Accretion Power}

Our model fits give a total accretion power of
$P_{\rm acc}=L_{\rm disk}/\eta_{\rm disk} = 1.9\times10^{47}$\ erg s$^{-1}$.  If the
bolometric isotropic equivalent luminosity from the 2010 June 17 flare
is $L_{\rm iso}=10^{48}$\ erg s$^{-1}$ \citep{tanaka11} then the radiative
efficiency of the flare is
\begin{eqnarray} 
\label{energy_constraint1}
\eta_j <
\frac{L_{\rm iso}}{2\Gamma^2(P_{j,e}+P_{j,B})} \approx 0.7 
\end{eqnarray}
\citep[][where the factor of 2 takes into account the two-sided
jet]{finke08_ssc,sikora09,tanaka11}, which implies a highly radiatively
efficient jet. The estimate for the total jet power, $P_j = P_{j,e}+P_{j,B}+P_{j,p}$, is a lower
limit because it does not include a contribution from protons in the jet ($P_{j,p}$),
which are likely to be present \citep[e.g.,][]{sikora00,sikora09}.  The jet power contributes a fraction of the
total accretion power of
\begin{eqnarray}
\label{energy_constraint2}
\frac{P_{j,e}+P_{j,B}}{P_{\rm acc}} = 2.3\times10^{-3}\ , 
\end{eqnarray}
although again note that this is a lower limit due to the uncertainty
of the proton content.  In fact, requiring that $P_j/P_{\rm acc}<1$ gives
a constraint on the power in protons in the jet $P_{j,p} \la 440
P_{j,e}$.  A low $P_j/P_{\rm acc}$ is at odds with the conclusions of
\citet{tanaka11} who estimate a much higher $P_j/P_{\rm acc}$.  The
difference is due to their assuming a smaller $\Gamma$ and $\eta_j$
than our derived values.

It is interesting to explore the possibility that the flare occurs
inside the BLR, and the break in the LAT spectrum is due to
$\gamma\gamma$ absorption of $\gamma$-rays with \ion{He}{2} Ly photons (continuum and lines)
\citep{poutanen10,stern11}.  Following \citet{tanaka11}, we find that
\begin{eqnarray}
\label{Lyman}
L_{\rm He\,II Ly} \cong 0.1 L_{\rm H\,I Ly\alpha} \cong 4.5 L_{\rm H\beta}  \cong 2.2\times10^{44} \,\rm erg\,\rm s^{-1}
\end{eqnarray}
where we have used the value for $L_{\rm H\beta}$ discussed above.  
Assuming the typical radius for the \ion{He}{2} emission is at a radius 
$R_{\rm He\,II} \cong 0.5 R_{\rm BLR}$, i.e., at $10^{17}$\ cm, we find that 
the spectral break one expects from $\gamma\gamma$ absorption with 
\ion{He}{2} Ly photons is
\begin{eqnarray}
\Delta\Gamma \sim \frac{\tau_T(5\ \rm GeV)}{4} 
\cong \frac{ \sigma_T L_{\rm He\,II Ly} }{ 16\pi c\ E_{\rm He\,II Ly}\ R_{\rm He\,II} } 
\cong 1.2\ , 
\end{eqnarray}
significantly larger than the $\Delta\Gamma\cong0.5$ found in the LAT
spectrum for 4C~+21.35.  The uncertainty in broad emission line
luminosities seems to make this approximation a rough estimate. We
note that a disk wind model for the BLR
\citep{murray95,chiang96,murray96} would lower the $\gamma\gamma$
opacity of the BLR, relative to a spherical shell geometry. We tested this possibility, however, and found that the $\gamma\gamma$
opacity remains extremely high, so it is still highly unlikely that
MAGIC-detected $\gamma$-ray photons could escape the BLR.

\begin{table*}
\footnotesize
\begin{center}
\caption{Model Parameters for the SED Shown in Figure~\ref{SED}\label{table_fit}. A black hole mass of $6\times\,10^8\ M_{\odot}$
  was considered.}
\begin{tabular}{lcccc}
\hline
Parameter & Symbol & 2010 June 17 & 2010 April 29 & Quiescent State \\
\hline
Gravitational radius  (cm)            &    $R_g$       & $8.8\times10^{13}$ & $8.8\times10^{13}$ & $8.8\times10^{13}$ \\
Eddington luminosity  (erg s$^{-1}$)         &   $L_{\rm Edd}$     & $7.8\times10^{46}$ & $7.8\times10^{46}$ & $7.8\times10^{46}$ \\
Disk Eddington ratio                  & $L_{\rm disk}/L_{\rm Edd}$ & 0.2 & 0.2 & 0.2 \\
Disk accretion efficiency             &   $\eta_{\rm disk}$       & 1/12 & 1/12 & 1/12 \\
Inner disk radius      ($R_g$)        &  $R_{\rm in}$ & 3 & 3 & 3 \\
Outer disk radius      ($R_g$)        &  $R_{\rm out}$ & 3$\times\,10^{4}$ & 3$\times\,10^{4}$ & 3$\times\,10^{4}$ \\
Bulk Lorentz factor                   & $\Gamma$  & 40 & 40 & 40	  \\
Doppler factor & $\delta_{\rm D}$      & 80 & 80 & 80    \\
Magnetic field & $B$  (G)            & 0.7  & 0.7  & 0.7 \\
Variability timescale & $t_v$ (s)     & 6$\times$$10^2$ & 6$\times$$10^2$ & 6$\times$$10^2$ \\
Comoving radius of blob & $R^{\prime}_b$ (cm) & 1.0$\times$10$^{15}$ & 1.0$\times$10$^{15}$ & 1.0$\times$10$^{15}$ \\
Jet height (cm) & $r$ & $8.8\times10^{18}$ & $8.8\times10^{18}$ & $8.8\times10^{18}$ \\
Low-energy electron spectral index & $p_1$       & 2.0 & 2.0 & 2.0     \\
High-energy electron spectral index  & $p_2$       & 3.5 & 3.5 & 3.5	 \\
Minimum electron Lorentz factor & $\gamma^{\prime}_{\rm min}$  & $1.0$ & $1.0$ & $1.0$ \\
Break electron Lorentz factor & $\gamma^{\prime}_{\rm brk}$ & $1.0\times10^3$ & $6.0\times10^2$ & $26$ \\
Maximum electron Lorentz factor & $\gamma^{\prime}_{\rm max}$  & $4.0\times10^4$ & $2.0\times10^4$ & $4.0\times10^4$  \\
Dust torus luminosity (erg s$^{-1}$) & $L_{\rm dust}$ & $5.5\times10^{45}$ & $5.5\times10^{45}$ & $5.5\times10^{45}$ \\
Dust torus temperature (K) & $T_{\rm dust}$ & $1.1\times10^3$ & $1.1\times10^3$ & $1.1\times10^3$ \\
Dust torus radius (cm) & $R_{\rm dust}$ & $1.8\times10^{19}$ & $1.8\times10^{19}$ & $1.8\times10^{19}$ \\
Jet power in magnetic field (erg s$^{-1}$) & $P_{j,B}$ & $5.9\times10^{42}$ & $5.9\times10^{42}$ & $5.9\times10^{42}$ \\
Jet power in electrons (erg s$^{-1}$) & $P_{j,e}$ & $4.3\times10^{44}$ & $4.0\times10^{44}$ & $1.9\times10^{44}$ \\
\hline
\end{tabular}
\end{center}
\end{table*}

\section{Conclusions}

4C $+$21.35 was detected at VHE by MAGIC on 2010 June 17 during a
period of high $\gamma$-ray activity detected by {\em Fermi}-LAT. The
relatively hard spectrum of the combined HE and VHE spectrum ($\Gamma$
= 2.7 $\pm$ 0.3), with no evidence of a cutoff, together with the
very rapid variability ($\sim$10 minutes) observed by MAGIC challenge
standard emission models. We presented multiwavelength observations of
the FSRQ 4C $+$21.35 collected from radio to VHE during
2009--2010. The first hint of a signal at VHE by MAGIC was found on
May 3, during a further period of $\gamma$-ray activity observed by
{\em Fermi}, suggesting a common origin for both the HE and VHE
emission during the 2010 April and June episodes.

During 2010 February--June only moderate flux variability was observed
in X-rays (a factor of $\sim$3), with the photon index changing in the
range 1.4--2.2 but with no correlation between flux and photon
index. A low variability amplitude was observed in UV in the same
period, suggesting that the UV is dominated by
the accretion disk emission that dilutes the jet emission. It is worth
noting that the peak of the UV emission was detected on June 18, but
the small increase observed makes it unlikely that the change of the
accretion rate is the main driver of the HE and VHE high activity
detected by {\em Fermi} and MAGIC. The optical light curve shows
variable flux density, but no dramatic increase of the activity. Two
optical peaks were observed on 2010 April 20 (MJD 55306) and 2010 June
30 (MJD 55377), close in time but not simultaneous with the $\gamma$-ray peaks.

Based on the 15 GHz MOJAVE data, there is no evidence for the ejection
of super-luminal knots associated with either of the flares in 2010
April and June.  However, \citet{marscher12} detected the
ejection of a superluminal knot with 43 GHz imaging at a time
somewhere between 2010 April 29 and June 3 (MJD 55315-55350), close in time
with the first 2010 $\gamma$-ray flare (see their Figure\ 3). We also noted that this knot could be associated with the
$\gamma$-ray outburst at around 2010 May 24 (MJD 55340). If the flare occurred
at the 43 GHz core, our model implies that the 43 GHz core is
  about 3 pc from the central BH. 

Based on our SED modeling (Section \ref{section:SEDmodel}), we reach
the following conclusions:
\begin{enumerate}
\item The $\gamma$-ray flares in 2010 April and June cannot  have originated from inside the BLR, at least not without invoking some
unusual particle transport mechanism \citep{dmt12,tavecchio12}.
\item There is some evidence for a rapidly-spinning prograde BH based on the optical emission.
\item The two flaring states and the quiescent state can be modeled by 
varying only the electron distribution for the source.
\end{enumerate}

The last result, modeling the source by varying only the
electron distribution, has also been found for the blazar
PKS\,0537$-$441 \citep{dammando13_0537}. This conclusion is much
stronger for PKS\,0537$-$441, since the optical continuum of PKS\,0537$-$441 is not disk-dominated, making its modeling more
constraining. Nonetheless, there are clearly sources for which a
change in the electron distribution is not sufficient to explain the
difference between flaring and quiescent states. For example, to model a strong optical-near infrared flare from
  PKS\,0208$-$512 with no counterpart in $\gamma$-rays required changing the magnetic
field strength \citep{chatterjee13}.

Rotation in polarization angles coincident with flares has been observed before in the blazars BL Lac \citep{marscher08}, PKS\,1510$-$089 \citep{marscher10,orienti13}, and 3C 279 \citep{abdo10_3c279}. They could be caused by a sudden realignment in the magnetic field due to shock compression, or a curved trajectory taken by the flaring region. A slight increase of the degree of optical polarization but no significant rotation of the polarization angle was observed at the time of the 2010 June HE and VHE flare.

The object 4C~+21.35 continues to challenge our understanding of
blazar emission mechanisms and the location of the emitting region.
Multi-wavelength observations have complemented previous LAT and MAGIC
observations to give a more complete picture for this source, although
many outstanding questions remain.

\acknowledgments
The {\em Fermi} LAT Collaboration acknowledges generous ongoing
support from a number of agencies and institutes that have supported
both the development and the operation of the LAT as well as
scientific data analysis.  These include the National Aeronautics and
Space Administration and the Department of Energy in the United
States, the Commissariat \`a l'Energie Atomique and the Centre
National de la Recherche Scientifique / Institut National de Physique
Nucl\'eaire et de Physique des Particules in France, the Agenzia
Spaziale Italiana and the Istituto Nazionale di Fisica Nucleare in
Italy, the Ministry of Education, Culture, Sports, Science and
Technology (MEXT), High Energy Accelerator Research Organization (KEK)
and Japan Aerospace Exploration Agency (JAXA) in Japan, and the
K.~A.~Wallenberg Foundation, the Swedish Research Council and the
Swedish National Space Board in Sweden. Additional support for science
analysis during the operations phase is gratefully acknowledged from
the Istituto Nazionale di Astrofisica in Italy and the Centre National
d'\'Etudes Spatiales in France. 

MAGIC Collaboration would like to thank the Instituto de Astrof\'{\i}sica de
Canarias for the excellent working conditions at the Observatorio del Roque de los Muchachos in La Palma.
The support of the German BMBF and MPG, the Italian INFN, the Swiss National Fund SNF, and the Spanish MICINN is
gratefully acknowledged. This work was also supported by the CPAN
CSD2007-00042 and MultiDark CSD2009-00064 projects of the Spanish Consolider-Ingenio 2010
programme, by grant 127740 of the Academy of Finland, by the DFG Cluster of Excellence ``Origin and Structure of the
Universe'', by the DFG Collaborative Research Centers SFB823/C4 and SFB876/C3,
and by the Polish MNiSzW grant 745/N-HESS-MAGIC/2010/0.

We thank the {\em Swift} team for making these observations possible,
the duty scientists, and science planners. This research has made use
of data from the MOJAVE database that is maintained by the MOJAVE team
(Lister et al. 2009, AJ, 137, 3718). The MOJAVE project is supported under 
NASA-Fermi grant 11-Fermi11-0019. The National Radio Astronomy Observatory is
a facility of the National Science Foundation operated under cooperative
agreement by Associated Universities, Inc. This work made use of the Swinburne
University of Technology software correlator (Deller et al. 2011, PASP, 123,
275), developed as part of the Australian Major National Research Facilities
Programme and operated under license. The OVRO 40-m monitoring program
is supported in part by NASA grants NNX08AW31G and NNX11A043G, and NSF
grants AST-0808050 and AST-1109911. This paper is partly based on
observations with the 100-m telescope of the MPIfR
(Max-Planck-Institut f\"ur Radioastronomie) at Effelsberg and the
Medicina telescope operated by INAF--Istituto di Radioastronomia. We
acknowledge A. Orlati, S. Righini, and the Enhanced Single-dish
Control System (ESCS) Development Team. We acknowledge financial
contribution from agreement ASI-INAF I/009/10/0.  The Submillimeter
Array is a joint project between the Smithsonian Astrophysical
Observatory and the Academia Sinica Institute of Astronomy and
Astrophysics and is funded by the Smithsonian Institution and the
Academia Sinica. Data from the Steward Observatory spectropolarimetric
monitoring project were used. This program is supported by Fermi Guest
Investigator grants NNX08AW56G, NNX09AU10G, and NNX12AO93G. The St. Petersburg
University team acknowledges support from Russian RFBR foundation, grants
12-02-00452 and 12-02-31193. The Abastumani team acknowledges financial
support of the project FR/638/6-320/12 by the Shota Rustaveli National Science
Foundation under contract 31/77. The Mets\"ahovi team acknowledges support
from the Academy of Finland to our observing projects (numbers 212656, 210338,
121148, and others). E.R. was partially supported by the Spanish MINECO
projects AYA2009-13036-C02-02 and AYA2012-38491-C02-01 and by the Generalitat
Valenciana project PROMETEO/2009/104, as well as by the COST MP0905 action
``Black Holes in a Violent Universe''.  Y.Y.K. was partly supported by the
Russian Foundation for Basic Research  (project 13-02-12103) and the Dynasty Foundation. We thank the anonymous referee for useful comments and suggestions.
J.F. would like to thank J.\ Steiner for useful discussions regarding the black hole spin of 4C $+$21.35.  F.D. would like to thank P.\ Smith for useful discussions about the polarimetric observations of 4C $+$21.35.

{\it Facilities:} \facility{{\em Fermi}}, \facility{MAGIC}, \facility{{\em Swift}}, \facility{ATOM}, \facility{KVA}, \facility{SMA},
\facility{Mets\"ahovi radio}, \facility{UMRAO}, \facility{OVRO:40m}, \facility{VLBA}, \facility{Medicina:32m}, \facility{St. Petersburg}, \facility{Effelsberg}, \facility{Abastumani}, \facility{Catalina}, \facility{Crimean}, \facility{SO:1.5m}.

\end{document}